%% file: sud11.tex
\documentclass[11pt]{article}
\usepackage{authblk}
\usepackage{amsthm}
\theoremstyle{definition}
\usepackage{amsmath,amsfonts,amssymb,amsthm,latexsym,stmaryrd}
\usepackage[dvips]{graphics}
\usepackage[all]{xy}
\usepackage{epic}
\usepackage{eepic}
\usepackage{times}

\usepackage{verbatim}

\newtheorem{thm}{Theorem}

\newtheorem{lem}[thm]{Lemma}
\newtheorem{defi}[thm]{Definition}

\newtheorem{cor}[thm]{Corollary}

\theoremstyle{remark}
\newtheorem*{rem}{Remark}
\newtheorem*{exa}{\bf Example}

\newcommand{\ket}[1]{\mathop{\big|#1\big>}\nolimits}       
\newcommand{\bra}[1]{\mathop{\big<#1\,\big|}\nolimits}     
\newcommand{\kb}[2]{\big| #1\big\rangle\!\big\langle #2 \big|}
\newcommand{\Tr}[1]{\mathop{{\mathrm{Tr}}_{#1}}}              

\newcommand{\ahead}[2]{\genfrac{}{}{0pt}{}{#1}{#2}}
\newcommand{\nn}{\nonumber}

\newcommand{\dg}{\dagger}

\def\openone{\mathbb{I}}

\def\N{\mathcal{N}}

\def\D{\mathcal{D}}
\def\C{\mathcal{C}}
\def\A{\mathcal{A}}

\def\H{\mathcal{H}}
\def\C{\mathcal{C}}

\def\E{\mathcal{E}}
\def\F{\mathcal{F}}

\def\B{\mathcal{B}}
\def\K{\mathcal{K}}

\def\O{\mathcal{O}}
\def\dm{\mathcal{DM}}
\def\Dcd{{\check{\D}}}

\def\bk{\mathbf k}

\def\a{\alpha}
\def\b{\beta}
\def\d{\delta}

\def\v{\varphi}
\def\e{\varepsilon}

\def\o{\omega}
\def\Om{\Omega}

\def\s{\sigma}

\def\lam{\lambda}

\newcommand{\slnc}{\mathsf{sl}(d,\mathbb{C})}
\newcommand{\sltc}{\mathsf{sl}(2,\mathbb{C})}
\newcommand{\slthc}{\mathsf{sl}(3,\mathbb{C})}

\newcommand{\proj}[1]{|#1\rangle\langle#1|}
\newcommand{\boson}[1]{a_{#1}^{\dagger}a_{#1}}
\newcommand{\cA}{\mathbb{A}}

\newcommand{\id}{\operatorname{id}}
\newcommand{\smfrac}[2]{\mbox{$\frac{#1}{#2}$}}

\begin{document}

\bibliographystyle{plain}

\title{{\bf Quantum Communication in Rindler Spacetime}}

\author[1]{Kamil Br\'adler}
\author[1,2]{Patrick Hayden}
\author[1]{Prakash Panangaden}
\affil[1]{School of Computer Science, McGill University, Montreal, Quebec, Canada}
\affil[2]{Perimeter Institute for Theoretical Physics, Waterloo, Ontario, Canada}

\date{18 September 2011}

\maketitle

\begin{abstract}
A state that an inertial observer in Minkowski space perceives to be
the vacuum will appear to an accelerating observer to be a thermal
bath of radiation. We study the impact of this Davies-Fulling-Unruh
noise on communication, particularly quantum communication from an
inertial sender to an accelerating observer and private
communication between two inertial observers in the presence of an
accelerating eavesdropper. In both cases, we establish compact,
tractable formulas for the associated communication capacities
assuming encodings that allow a single excitation in one of a fixed
number of modes per use of the communications channel. Our
contributions include a rigorous presentation of the general theory
of the private quantum capacity as well as a detailed analysis of
the structure of these channels, including their group-theoretic
properties and a proof that they are conjugate degradable.
Connections between the Unruh channel and optical amplifiers are
also discussed.
\end{abstract}


\parskip .75ex

\section{Introduction}
A well-known feature of quantum field theory in curved spacetimes
is the creation of particles from a vacuum~\cite{parker68}, which
points to a fundamental ambiguity: the notion of particle is not an
absolute one in the absence of Poincar\'e invariance.  Even in flat
spacetimes one has the Davies-Fulling-Unruh
effect~\cite{fulling73,davies75,unruh76,unruh84} whereby a uniformly
accelerating observer in Minkowski space detects a thermal bath of
radiation in a state that an inertial observer perceives as a
vacuum.  This phenomenon is symptomatic of a nonuniqueness in the
definition of the vacuum state of quantum field theory in curved
spacetimes in the absence of some canonical symmetry consideration
that allows one to choose a preferred vacuum state.

In quantum information theory, on the other hand, one typically
treats the notion of particle as canonical and concepts like ``pure
state'' and ``mixed state'' are taken to have absolute meaning.  In
the present work, we examine the consequences for quantum
information theory of this ambiguity in the definition of vacuum
(and particle) states.
Specifically, we study optimal communications strategies in the face
of these relativistic difficulties, building on earlier studies of
how relativistic effects impact entanglement manipulation and
quantum communications
strategies~\cite{AlsingMilburn,Peres04,gingadami,cabremb05,doukas2010entanglement,fs,datta2009discord}.
While most such work studied the degradation caused when protocols
not designed for relativistic situations are employed in situations
where relativistic effects are significant, our approach will be to
design protocols specifically with relativistic effects in mind, in
the spirit of~\cite{Kent99,CW03,qubitunruh,cliche2010}.

We focus on two scenarios. In the first, an inertial observer, Alice,
attempts to send quantum information to an accelerating receiver, Bob, by
physically transmitting scalar ``photons'' of chosen modes. Owing to
the thermal noise perceived by the receiver, quantum error
correcting codes are required to protect the quantum information.

The second scenario is more elaborate. Two inertial observers --
again call them Alice and Bob -- communicate by exchanging
scalar ``photons'' of chosen modes, while an accelerating observer
-- traditionally called Eve -- attempts to eavesdrop or
wiretap their communication channel.  This time, it is Eve who
detects thermal noise and therefore cannot perfectly decode the
communications between Alice and Bob, thus allowing the possibility of
private communication between them.  Of course, we are not
proposing this as a practical scheme for cryptography but, rather,
as an exploration of the impact of relativistic quantum field theory
on quantum information theory.

The concept of private capacity in the classical setting is due to
Maurer~\cite{maurer1994strong} and independently Ahlswede and
Csiszar~\cite{ahlswede1993common}.  The private capacity of a
quantum channel was first studied by Cai \emph{et al.} and
Devetak~\cite{CWY:privacy,Devetak05}. These capacities measure the
optimal rate at which Alice can transmit classical bits to Bob that
remain secret from Eve, in the limit of many uses of the channel. In
the present paper we introduce the private \emph{quantum} capacity
of a quantum channel, which measures the usefulness of the channel
for sending private quantum mechanical data (qubits) instead of
bits.

The standard approach to quantum field theory in flat spacetime is
to decompose the field into ``positive'' and ``negative'' frequency
modes as defined by the Fourier transform.  One then defines
creation and annihilation operators that correspond to these modes
and the vacuum state is defined to be the state killed by all the
annihilation operators.  The Poincar\'e invariance of Minkowski
spacetimes means that the vacuum state is the unique state that is
invariant under the action of the Poincar\'e group.  In Rindler
space, it is natural for the accelerating observer to use his or her
own timelike Killing field to define the notion of positive and
negative frequency.  This means that there will be a mismatch
between Alice's notion of vacuum state and that of the accelerating
observer. The transformation between the creation and annihilation
operators of the different (and inequivalent) quantum field
theories is given by a linear map, called a \emph{Bogoliubov
  transformation}, between the creation and annihilation operators of the
two quantum field theories.

The explicit form of the Bogoliubov transformation is well known
and we use it to define a \emph{channel} which we call the Unruh
channel.  In quantum information theory, a channel is simply any
physically realizable transformation of a quantum state.  The idea
is that the process of transmission may introduce noise and loss of
information.  Thus, an initially pure quantum state may become
mixed.

\begin{figure}[t]
\begin{center}
    \resizebox{13.5cm}{7cm}{\includegraphics{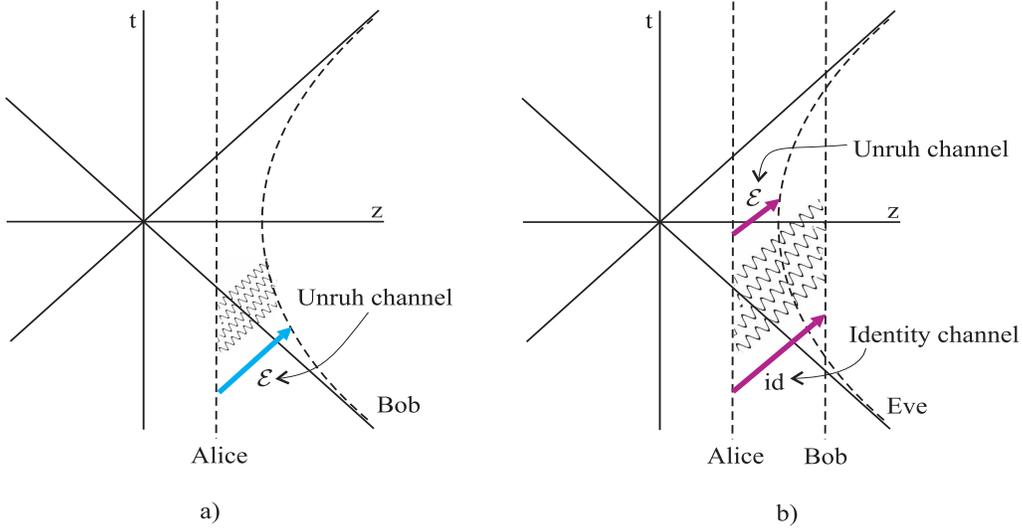}}
    \caption{Spacetime diagrams for the two communication scenarios. (a) Alice is an inertial
    observer try to send quantum information to the uniformly accelerated Bob. The wavy lines
    indicate transmission via wave packets and the $d$-rail qudit encoding. (b) In the second
    diagram, Alice and the intended receiver, Bob, are both inertial observers. In our idealized scenario, they are assumed to share a noiseless quantum channel. A uniformly accelerated eavesdropper, Eve, attempts to wiretap Alice's message to Bob.}
    \label{fig:spacetime}
\end{center}
\end{figure}

In the Unruh channel, Alice prepares some state in her chosen
$d$-dimensional space encoded in terms of Minkowski modes.  An
accelerating observer (Bob or Eve depending on the scenario)
intercepts this, but using an apparatus that detects excitations of
the quantum field defined according to the prescription of the
Rindler quantum field theory. So the state that she detects will be
described by some infinite-dimensional density matrix.  A detailed
analysis of this density matrix makes it possible to extract
quantitative information about the private and quantum capacities.
We evaluate both the quantum capacity from Alice to an accelerating
Bob \emph{and} the private capacity for inertial Alice and Bob
trying to exchange quantum information while simultaneously
confounding an accelerating eavesdropper. Figure \ref{fig:spacetime}
contains spacetime diagrams illustrating the two communication
scenarios.

Both quantities exhibit surprising behavior.  The quantum capacity,
the optimal rate at which a sender can transmit qubits to a receiver
through some noisy channel, usually exhibits a threshold behavior;
channels below some quality threshold have quantum capacity exactly
zero.  For the Unruh channels, however, we find that the quantum
capacity is strictly positive for all accelerations, reaching zero
only in the limit of infinite acceleration.  It is therefore always
possible to transmit quantum data to an accelerating receiver
provided the sender is not behind the receiver's horizon.  Careful
choices of encoding can therefore eliminate the degradation in
fidelity known to occur if one uses a naive teleportation protocol
to communicate with an accelerating receiver~\cite{AlsingMilburn} (see also~\cite{schutzhold2005comment}).
In addition to characterizing quantum transmission to an
accelerating receiver, our analysis applies equally well to the
study of quantum data transmission through an optical amplifier,
which may well be its more important application.

The private quantum capacity is likewise positive for all nonzero
eavesdropper accelerations.  Thus, in principle, any eavesdropper
acceleration, no matter how small, can be exploited to safeguard
transmissions of quantum data between two inertial observers.
Curiously, the private quantum capacity has a simple formula when
the channel between the inertial observers is noiseless; the formula
reveals that in this case the private quantum capacity is exactly
equal to the entanglement-assisted quantum capacity to the
eavesdropper's environment, despite the absence of any entanglement
assistance in the problem.

\subsection{Structure of the paper} \label{subsec:structure}

Section \ref{subsec:qcap} reviews the definition of the quantum capacity and states the Lloyd-Shor-Devetak theorem, which provides the best known achievable rates for quantum data transmission over noisy channels. Section \ref{subsec:privcap} introduces the private quantum capacity and proves a capacity theorem in the case where the channel to the intended recipient is noiseless. Section \ref{subsec:Unruh} reviews the Unruh effect, which then allows for an analysis of the output density matrix of the Unruh channel in Section \ref{subsec:geomstruct}. Section \ref{sec:capcalc} is devoted to the explicit capacity calculations.

\subsection{Notation} \label{subsec:notation}

If $A$ and $B$ are two  Hilbert spaces, we write $AB \equiv A\otimes
B$ for their tensor product.  The Hilbert spaces on which linear
operators act will be denoted by a subscript.  For instance, we
write $\v_{AB}$ for a density operator on $AB$.  Partial traces will
be abbreviated by omitting subscripts, such as $\v_A \equiv
\Tr{B}\v_{AB}$.  We use a similar notation for pure states, e.g.\
$\ket{\psi}_{AB}\in AB$, while abbreviating $\psi_{AB} \equiv
\kb{\psi}{\psi}_{AB}$.  We will write $\id_A$ for the identity
channel acting on $A$. In general, the phrase \emph{quantum channel}
refers to a completely positive, trace-preserving linear map.
The symbol $\openone_A$ will be reserved for the identity matrix acting on
the Hilbert space $A$ and $\pi_A = \openone_A / \dim A$ for the maximally
mixed state on $A$. The identity with a superscript $\openone^{(k)}$ used in Sec.~\ref{subsec:geomstruct} and onwards acts on a ${d+k-1\choose k}$-dimensional Hilbert space. The symbol $\Phi$ will be reserved for maximally
entangled states and, in particular, $\ket{\Phi_{2^k}} = 2^{-k/2}
\sum_{j=1}^{2^k} \ket{j}\ket{j}$ will denote the maximally entangled
state on $k$ pairs of qubits.

The trace norm of an operator, $\|X\|_1$ is defined to be $\Tr{}|X|
= \Tr{}\sqrt{X^\dagger X}$.  The similarity of two density operators
$\v$ and $\psi$ can be measured by \emph{trace distance}
$\smfrac{1}{2} \| \v - \psi \|_1$, which is equal to the maximum
over all possible measurements of the variational distance between
the outcome probabilities for the two states.  The trace distance is
zero for identical states and one for perfectly distinguishable
states.

A complementary measure is the mixed state fidelity
\begin{equation} \label{eq:fidelity.defn}
  F(\v,\psi)
  = \left\| \sqrt{\v}\sqrt{\psi} \right\|_1^2
  = \left( \Tr{}\sqrt{\sqrt{\v}\psi\sqrt{\v}} \right)^2,
\end{equation}
defined such that when one of the states is pure, $F(\v,\psi) =
\Tr{}\v\psi$.  More generally, the fidelity is equal to one for
identical states and zero for perfectly distinguishable states.


For a density operator $\sigma_{AB}$, let $H(A)_\sigma$ be the von
Neumann entropy of $\sigma_A$.  The \emph{mutual information}
$I(A;B)_\sigma$ is $H(A)_\sigma + H(B)_\sigma - H(AB)_\sigma$ while
the \emph{coherent information} is $I(A\rangle B)_\sigma =
H(B)_\sigma - H(AB)_\sigma$. The latter quantity, as the negation of
the concave conditional entropy $H(A|B)_\sigma = H(AB)_\sigma - H(B)_\sigma$,
can be positive only when the state $\sigma$ is entangled.

For more information on the properties of quantum channels or the functions defined here,
we refer the reader to Nielsen and Chuang~\cite{Nielsen00}.

\section{Standard and Private Quantum Capacities}
\label{sec:capa}

The objective of the paper will be to evaluate two quantities
characterizing communication over the qudit Unruh channels: their
quantum capacity and private quantum capacity. While the quantum
capacity of a quantum channel has been studied in great
detail~\cite{bns:qcapconverse,lloyd1997capacity,Shor02,Devetak05,Hayden08,HSW:qcap,HLW:qcap,klesse:qcap}, the private quantum capacity of a wiretap channel
has not. A recent paper by Brand\~{a}o and Oppenheim does, however, consider the very interesting and somewhat related problem of using a fixed, shared quantum state supplemented by public communication to securely transmit quantum information~\cite{1004.3328}. After briefly introducing the quantum capacity we will
therefore develop the general theory of the private quantum
capacity, rigorously demonstrating results that were only briefly
sketched in~\cite{qubitunruh}.

\subsection{Quantum Capacity}
\label{subsec:qcap}

The ability of a quantum channel to transmit quantum information is
measured by its quantum capacity, the optimal rate at which qubits
can be reliably transmitted in the limit of many uses of the channel
and vanishing error.
There are many equivalent ways to define the quantum capacity~\cite{kw:variazioni}. Here we use
a version which focuses on the transmission of halves of maximally entangled states across
the noisy channel.
Recall that $\ket{\Phi_{2^k}}$ represents the maximally
 entangled state on $k$ pairs of qubits.
\begin{defi} \label{defi:quantum.code}
An $(n,k,\delta)$ \emph{entanglement transmission code} from Alice
to Bob consists of an encoding channel $\A$ taking a $k$-qubit
system $R'$ into the input of $\N^{\otimes n}$ and a decoding
channel $\B$ taking the output of $\N^{\otimes n}$ to a $k$-qubit
system $C \cong R'$ satisfying
\begin{equation}
 \left\|
    (\id \otimes \B \circ \N^{\otimes n} \circ \A)
    ( \Phi_{2^k} ) -
     \Phi_{2^k}
 \right\|_1 \leq \delta.
\end{equation}
A rate $Q$ is an \emph{achievable rate} for entanglement
transmission if for all $\delta > 0$ and sufficiently large $n$
there exist $(n,\lfloor nQ \rfloor, \delta)$ entanglement
transmission codes.  The \emph{quantum capacity} $Q(\N)$ is the
supremum of all the achievable rates.
\end{defi}

In any capacity problem, the objective is to understand the
structure of the optimal codes. Doing so normally results in a
theorem characterizing the capacity in terms of simple entropic
functions optimized over a single use of the channel, a so-called
``single-letter formula.'' In general, the structure of the optimal
codes is still unknown for the quantum capacity problem. We will see
below, however, that they can be characterized in the case of qudit
Unruh channels.

The following theorem gives the best known general achievable rates
for the quantum capacity problem in terms of the coherent information, as
defined in the previous section.
\begin{thm}[Lloyd-Shor-Devetak~\cite{lloyd1997capacity,Shor02,Devetak05}] \label{thm:lsd}
Let $\ket{\psi}_{A'A}$ be a pure state, $\N$ a quantum channel from
$A$ to $B$ and define $\rho = (\id_{A'} \otimes \N)(\psi)$.  The
quantum capacity $Q(\N)$ of $\N$ is at least $I(A'\rangle B)_\rho$.
\end{thm}
Note that while $I(A'\rangle B)_\rho$ is expressed as a function of the pure state $\ket{\psi}^{A'A}$,  it is left invariant by unitary transformations of the $A'$ system and can, therefore, equally well be written as a function of the reduced density operator $\psi^A$. We will make use of that invariance in our calculations.

\subsection{Private Quantum Capacity: General Case}
\label{subsec:privcap}

\newcommand{\Ecb}{E_c}
\newcommand{\Ec}{E_c}
\newcommand{\Bcb}{{B_c}}
\newcommand{\Bc}{{B_c}}
\newcommand{\Cc}{{C_c}}

The private quantum capacity is the optimal rate at which a sender
(Alice) can send qubits to a receiver (Bob) while simultaneously
ensuring that those qubits remain encrypted from the eavesdropper's
(Eve's) point of view.  At first glance, this would not seem to be a
very interesting concept.  The impossibility of measuring quantum
information without disturbing it would seem to ensure that
successful transmission of quantum information would make it
automatically private.  One can imagine a passive eavesdropper,
however, who \emph{could} have nontrivial access to the qubits
should she choose to exercise it.  The setting we will ultimately be
primarily concerned with here is a relativistic version of that
passive eavesdropper, in particular, the case in which the
eavesdropper is uniformly accelerated.

\begin{figure}[t]
\begin{center}
\input{sud_wiretap1.eepic}
\end{center}
\caption{Another scenario in which the wiretap framework applies.
Alice sends quantum data to Bob through two separate channels, two
different fiber optic links, for example.  Eve potentially has access
to one of the links and Alice wants to ensure that should Eve try to
eavesdrop that she will not learn anything about the transmission.
The map $\N^{\otimes n}$ would represent all the noise experienced by both transmission lines
while Eve's channel $\E^{\otimes n}$ would describe the output of the transmission line entering her domain, not including any further noise it experiences before finally ending in Bob's laboratory. The dotted lines indicate that $\N^{\otimes n}$ and $\E^{\otimes n}$ should not be composed; each is a complete description of the noise experienced by Bob and Eve, respectively.}  \label{fig:wiretap}
\end{figure}
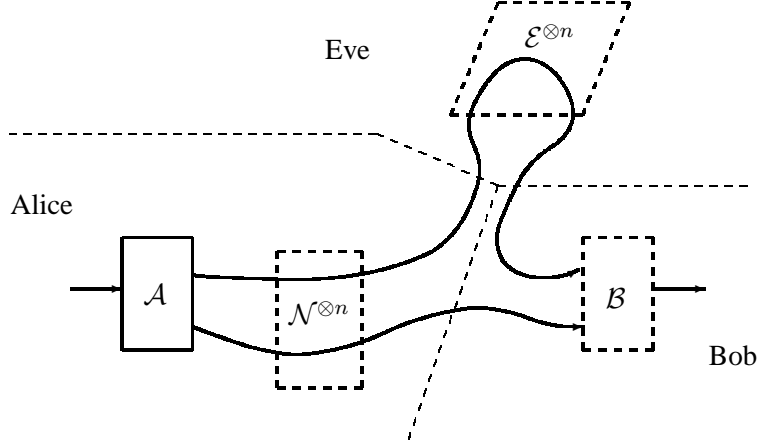

\begin{defi} \label{defi:wiretap.channel}
A \textbf{quantum wiretap channel} consists of a pair of quantum channels \\ $(\N_{A\rightarrow B},\E_{A\rightarrow E})$ taking the density operators on $A$ to those on $B$ and $E$, respectively.
\end{defi}
$\N$ should be interpreted as the channel from Alice to Bob and $\E$
the channel from Alice to Eve.  Let $U_\N : A \rightarrow B \otimes
{\Bc}$ and $U_\E : A \rightarrow E \otimes E_c$ be isometric
extensions of the channels $\N$ and $\E$.  In particular, $\N(\cdot)
= \Tr{{\Bc}} U_\N \cdot U_\N^\dagger$ and $\E(\cdot) = \Tr{E_c} U_\E
\cdot U_\E^\dagger$.  In many circumstances, $\E$ will be a degraded
version of the ``environment'' of the Alice-Bob channel, meaning
that there exists a channel $\D$ such that $\E(\cdot) = \D \circ
\Tr{B} U_\N \cdot U_\N^\dg$.  For the uniformly accelerated
eavesdropper, however, this needn't be the case so we don't require
\emph{a priori} that there be a particular relationship between $\N$
and $\E$.  Another relevant example is illustrated in Figure
\ref{fig:wiretap}.

Recall that  $\pi_{2^k} = \openone/2^k$,
the maximally mixed state on $k$ qubits.

\begin{defi} \label{defi:privacy.code}
An $(n,k,\delta,\epsilon)$ \textbf{private entanglement transmission
code} from Alice to Bob consists of an encoding channel $\A$ taking
a $k$-qubit system $R'$ into the input of $\N^{\otimes n}$ and a
decoding channel $\B$ taking the output of $\N^{\otimes n}$ to a
$k$-qubit system $C \cong R'$ satisfying
\begin{enumerate}
\item Transmission:
$
 \left\|
    (\id \otimes \B \circ \N^{\otimes n} \circ \A)
    ( \Phi_{2^k} ) -
     \Phi_{2^k}
 \right\|_1 \leq \delta.
 $
\item Privacy:
$
 \left\|
    (\id \otimes \E^{\otimes n} \circ \A)
    ( \Phi_{2^k} )   -
   \pi_{2^k} \, \otimes (\E^{\otimes n} \circ \A) (
   \pi_{2^k} )
 \right\|_1\leq \epsilon.
 $
\end{enumerate}
A rate $Q$ is an \textbf{achievable rate} for private entanglement
transmission if for all $\delta, \epsilon > 0$ and sufficiently
large $n$ there exist $(n,\lfloor nQ \rfloor, \delta, \epsilon)$
private entanglement transmission codes.  The \textbf{private quantum
capacity} $Q_p(\N,\E)$ is the supremum of all the achievable rates.
\end{defi}

The transmission criterion states that halves of EPR pairs encoded
by $\A$, sent through the channel and then decoded by $\B$ will be
preserved by the communications system with high fidelity.
Alternatively, one could ask that arbitrary pure states or even
arbitrary states entangled with a reference sent through $\B \circ
\N^{\otimes n} \circ \A$ be preserved with high fidelity.  The
different definitions are equivalent for the standard quantum
capacity $Q(\N) = Q_p(\N,\Tr{})$, which is defined with no privacy
requirement~\cite{kw:variazioni}.  The equivalence extends straightforwardly to
the private quantum capacity.

The privacy condition can also be written in a slightly more
indirect but illustrative way.  If $\Psi_{RE^n} =   (\id_R \otimes
\E^{\otimes n} \circ \A)( \Phi_{2^k} )$, then the condition states
that
\begin{equation} \label{eqn:decoupling}
 \left\|
    \Psi_{RE^n} - \Psi_{R} \otimes \Psi_{E^n}
 \right\|_1\leq \epsilon.
\end{equation}
In words, the channel $\E^{\otimes n} \circ \A$ should destroy all
correlations with $R$ for the input maximally entangled state
$\Phi_{2^k}$.

Let $\E_c(\cdot) = \Tr{E} U_\E \cdot U_\E^\dagger$ be the channel
from Alice to the environment of the channel to Eve.  The output of
$\E_c$ contains data that Eve is incapable of intercepting, which
explains its appearance in our main capacity theorem:
\begin{thm}[Private quantum capacity] \label{thm:qpriv}
The private quantum capacity $Q_p(\id,\E)$ when the channel from
Alice to Bob is noiseless is given by the formula $\max
\smfrac{1}{2} I(A';E_c)_\rho$, where the maximization is over all
pure states $\ket{\psi}_{A'A}$ and $\rho = (\id \otimes
\E_c)(\psi)$.
\end{thm}

Because the mutual information is equal to zero only for product
states, $Q_p(\id,\E)$ is zero only when $\E_c$ is the constant
channel or, equivalently, $\E$ is the identity.  In particular, it is
not necessary for $\E_c$ to have nonzero quantum capacity in order
for $Q_p(\id,\E)$ to be positive.  The fact that the optimization is
over input states to a single copy of $\E$ is notable: the number of
such ``single-letter'' results in quantum Shannon theory is very
limited.  No single-letter formulas are known for the classical or
quantum capacities of general quantum channels, for example.

Despite the absence here of any entanglement assistance, the theorem
implies that $Q_p(\id,\E)$ is exactly equal to the
entanglement-assisted quantum capacity of $\E_c$, usually written
$Q_E(\E_c)$, by virtue of the fact that their formulas
match~\cite{BSST02}.  Why they should be the same is, however,
something of a mystery.

\begin{figure}
\begin{center}
\input{sud_codestructure.eepic}
\end{center}
\caption{Structure of a quantum privacy code.  $V_\A$, $U_\N$,
$U_\E$, $V_\B$ and $V_\D$ are isometric extensions of $\A$, $\N$,
$\E$, $\B$ and $\D$, respectively.  The initial state is maximally
entangled between $R$ and $R'$ and, because all the transformations
are isometries, the state remains pure as time increases from left
to right.  Registers meeting at a vertex on the right hand side
of the diagram are generically correlated while those not meeting
will be product.  (a) The transmission condition states that using
only the output of $\N^{\otimes n}$, Bob should be able to produce
the purification of the reference entanglement.  This implies, in
particular, that the reduced state on $R \otimes F$ is nearly
product, a fact used in the converse proof.  (b) The privacy
condition requires that the state on $R \otimes E^n$ be nearly
product.  That is equivalent to the existence of a decoding channel
$\D$ (with isometric extension $V_\D$) acting on $\Ecb^n \otimes F$
whose output approximates a purification of the reference
entanglement $R$.  The code construction demonstrates the existence
of such a $\D$. Note that while both $\B$ and $\D$ decode the same
quantum information, they cannot be applied simultaneously so the no-cloning
theorem is not violated.} \label{fig:code.structure}
\end{figure}
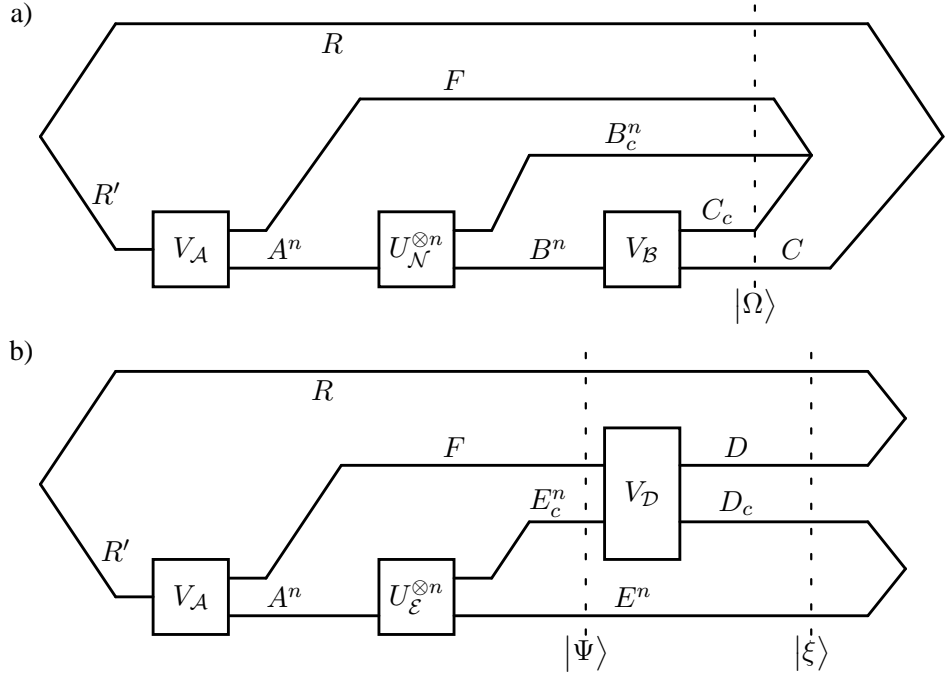

We will break the proof of Theorem \ref{thm:qpriv} into two parts,
the achievability of the claimed rate and then a converse showing
that it is impossible to do better.  The strategy is illustrated in
Figure \ref{fig:code.structure}.

The achievability part relies on the following simple lemma:
\begin{lem} \label{lem:duality.decoupling}
Let $\ket{\rho}_{AB}$ be a bipartite pure state and $\ket{\psi}_{A}$ a
pure state of $A$.  If $\| \rho_{A} - \psi_{A} \|_1 \leq \kappa$
then there exists a pure state $\ket{\omega}_B$ such that $\|
\rho_{AB} - \psi_{A} \otimes \omega_B \|_1 \leq 2 \sqrt{\kappa}$.
\end{lem}
\begin{proof}
Recall that, for all states $\phi$ and $\tau$, the mixed state
fidelity function satisfies
\begin{eqnarray}
    F(\phi,\tau)
        &\geq& 1 - \| \phi - \tau \|_1 \label{eqn:fid.trace} \\
    \quad \mbox{and} \quad
    \| \phi - \tau \|_1
        &\leq& 2 \sqrt{1-F(\phi,\tau)}.  \label{eqn:trace.fid}
\end{eqnarray}
(See, for example, \cite{Nielsen00}.) So, by the hypothesis of the lemma, $F(\rho_{A},\psi_{A}) \geq 1 -
\kappa$.  But by Uhlmann's theorem~\cite{uhlmann:fid,jozsa:fid},
\begin{equation*}
    F(\rho_{A},\psi_{A})
    = \max_{\ket{\omega}_B} \left| \bra{\rho}_{AB} \ket{\psi}_{A} \ket{\omega}_B \right|^2.
\end{equation*}
which completes the proof when combined with (\ref{eqn:trace.fid}).
\end{proof}

We will also need the following variant of the Lloyd-Shor-Devetak theorem:
\begin{thm} \label{thm:lsd++}
Let $\ket{\psi}_{A'A}$ be a pure state, $\N_j$ a quantum channel
from $A$ to $B_j$ for $1 \leq j \leq k$ and $\rho_j = (\id_{A'}
\otimes \N_j)(\psi)$. There is a single encoding $\A$ that will
achieve entanglement transmission for all $j$ at the rate
\begin{equation}
    \min_{1 \leq j \leq k} I(A' \rangle B_j)_{\rho_j}.
\end{equation}
\end{thm}
\begin{proof}
This is a special case of Theorem IV.3 of \cite{BBN08} except for
the fact that the theorem in question assumes that the output spaces
$B_j$ are all identical.  To apply the theorem, it therefore suffices
to set $B = \oplus_{j=1}^k B_j$ and compose each channel $\N_j$ with
the embedding of $B_j$ into $B$.

Alternatively, one can observe that the encodings used in
\cite{Hayden08} to achieve entanglement transmission at the coherent
information rate depend only on $\psi$ and not  on the channels
themselves.  The analysis therein demonstrates that for sufficiently
large $n$,  a random encoding succeeds for a given channel with high
probability.  Random encodings will therefore succeed for any finite
number of channels simultaneously again with high probability.
\end{proof}

\begin{proof} \emph{Achievability part of Theorem \ref{thm:qpriv}.}
Let $V_\A$ be an isometric extension of $\A$ with output on $A^n F$.
The privacy condition applied to $E^n$ is actually equivalent to
entanglement transmission to $F\Ecb^n$.  To show achievability, it
suffices to show that entanglement transmission implies privacy.
Indeed, suppose that there exists a ``decoding'' channel $\D$ from
$F\Ecb^n$ to a space of $k$ qubits on $D$ such that
\begin{equation*}
 \left\|
    (\id \otimes \D \circ \E_c^{\otimes n} )
    ( (\openone \otimes V_\A) \Phi_{2^k} (\openone \otimes V_\A^\dagger)) -
     \Phi_{2^k}
 \right\|_1 \leq \kappa.
\end{equation*}
Let $V_\D : F\Ecb^n \rightarrow DD_c$ be an isometric extension for
$\D$.  Call $\ket{\xi}_{RDD_cE^n}$ the purification of
    $(\id \otimes \D \circ \E_c^{\otimes n} )
    ( (\openone \otimes V_\A) \Phi_{2^k} (\openone \otimes V_\A^\dagger))$.
By Lemma \ref{lem:duality.decoupling} and as illustrated in Figure
\ref{fig:code.structure}b, there exists a pure state
$\ket{\omega}_{D'E^n}$ such that
\begin{equation}
    \left\| \xi_{RDD_cE^n} - (\Phi_{2^k})_{RD} \otimes \omega_{D_cE^n} \right\|_1
    \leq 2\sqrt{\kappa}.
\end{equation}
By the monotonicity of the trace distance under the partial trace,
this implies that
\begin{equation}
    \left\| \xi_{RE^n} - (\pi_{2^k})_{R} \otimes \omega_{E^n} \right\|_1
    \leq 2\sqrt{\kappa},
\end{equation}
which is nothing other than Eq.~(\ref{eqn:decoupling}) for $\epsilon
= 2\sqrt{\kappa}$.

It is therefore sufficient to find codes that simultaneously perform
entanglement transmission to $B^n$ and to $F\Ecb^n$, in the first
case for the channel $\id_{A^n\rightarrow B^n} \otimes \Tr{F}$
which traces over $F$ and in the second case for the
channel $\E_c^{\otimes n} \otimes
\id_F$ whose output combines $F$ with Eve's
complementary channel.  Applying Theorem \ref{thm:lsd++} to these
channels using the input state $\ket{\varphi} = \ket{\psi}^{\otimes
n}_{AA'} \otimes \ket{\Phi}_{FF'}$ provides the following pair of
conditions sufficient for simultaneous entanglement transmission:
\begin{eqnarray}
  nQ
    &<& I(A'F'\rangle B^n)_{\psi^{\otimes n} \otimes \Phi_{F'}}  \\
    &=&  H(B^n)_{\psi^{\otimes n}} - H(A'F'B^n)_{\psi^{\otimes n} \otimes \Phi_{F'}} \\
    &=& nH(A')_\psi - \log \dim F  \\
\mbox{and} \quad
nQ
    &<& I(A'F'\rangle F\Ecb^n)_{(\id_{A'F} \otimes \E_c^{\otimes n})(\varphi)} \\
    &=& n I(A'\rangle E_c)_\rho + \log \dim F.
\end{eqnarray}
(The expressions use the slight abuse of notation that $\psi_{A'B} = \id_{A \rightarrow B}(\psi_{A'A})$.) The simplifications rely only on the facts that the entropy of a
product state is the sum of the entropies of the individual factors
and that for any pure state $\ket{\o}_{XY}$, the nonzero eigenvalues
of $\o_X$ and $\o_Y$ are the same so that $H(\o_X) = H(\o_Y)$.

Choosing $\dim F =2^{nf}$ allows us to rewrite these conditions as
\begin{equation}
Q < H(A')_\psi - f \quad\mbox{and}\quad Q < I(A'\rangle E_c)_\rho +
f.
\end{equation}
The constraints have intuitive interpretations: the first is the
noiseless rate to Bob through $\id_A$ reduced by the rate at which
qubits are lost to $F$, while the second is the standard coherent
information rate for $\E_c$ augmented by a noiseless channel to $F$.
$Q$ is maximized subject to these constraints when $H(A')_\psi - f =
I(A'\rangle E_c)_\rho + f$.  Using the fact that $H(A)_\psi =
H(A')_\rho$ and purifying $\rho$ to $\ket{\rho}_{A'EE_c}$, this
equation can be written as
\begin{eqnarray}
  f
    &=& \smfrac{1}{2} \left[ H(A')_\psi - I(A'\rangle E_c)_\rho \right] \\
    &=& \smfrac{1}{2} \left[ H(A')_\rho - H(\Ec)_\rho + H(A'E_c)_\rho \right] \\
    &=& \smfrac{1}{2} \left[ H(A')_\rho - H(A'E)_\rho + H(E)_\rho \right] \\
    &=& \smfrac{1}{2} I(A';E)_\rho.
\end{eqnarray}
Therefore, the rate $Q$ is achievable
provided
\begin{eqnarray}
  Q
    &<& H(A')_\rho - \smfrac{1}{2}I(A';E)_\rho \\
    &=& H(A')_\rho - \smfrac{1}{2} \left[ H(A')_\rho - H(A'E)_\rho + H(E)_\rho \right] \\
    &=& \smfrac{1}{2} \left[ H(A')_\rho + H(\Ec)_\rho - H(A'E_c)_\rho \right] \\
    &=& \smfrac{1}{2}I(A';E_c)_\rho,
\end{eqnarray}
which is what we set out to prove.
\end{proof}
It wasn't essential that the channel from Alice to Bob be noiseless
until the entropic manipulations in the second half of the proof.
Stopping before that point provides the following achievable rates
in the general case:
\begin{cor} \label{cor:general.rate}
Let $(\N,\E)$ be a quantum wiretap channel.  For $\ket{\psi}_{A'A}$
any pure state, $\rho = (\id \otimes \N)(\psi)$ and $\tau = (\id
\otimes \E)(\psi)$, the following lower bound on the private quantum
capacity holds:
\begin{eqnarray}
Q_p(\N,\E)
    &\geq& \frac{1}{2} \left[ I(A'\rangle B)_\rho - I(A'\rangle
    E)_\tau \right].
\end{eqnarray}
\end{cor}

The proof of the converse to Theorem \ref{thm:qpriv} will rely on an
elegant inequality of Alicki and Fannes~\cite{AF03}:
\begin{lem} \label{lem:af.inequality}
Let $\rho_{AB}$ and $\sigma_{AB}$ be bipartite density operators on
finite dimensional systems and let $h_2(x) = - x \log x - (1-x) \log
(1-x)$. If $\| \rho_{AB} - \sigma_{AB} \|_1 \leq \e \leq 1/e$, then
\begin{equation}
    \left| H(A|B)_\rho - H(A|B)_\sigma \right|
    \leq 4 \epsilon \log \dim A + 2 \, h_2(\epsilon).
\end{equation}
\end{lem}
What is notable about the inequality is that the upper bound is
independent of the dimension of $B$.  In classical information
theory, a similar bound holds but for a trivial reason: if $\rho$ is
classical then $H(A|B)_\rho$ is an average of entropies of $A$
alone.  No such reduction exists in the quantum case, but $H(A|B)$
nonetheless behaves as if there were in this sense.

We will also need the other half of the equivalence between privacy and entanglement transmission.  Specifically, privacy implies entanglement transmission in the following sense:
\begin{lem} \label{lem:privacy.entanglement}
Let  $V : A \rightarrow BB_c$ be an isometric extension of some channel $\N$ from $A$ to $B$.  Fixing a Hilbert space $R$ satisfying $|R| \leq |A|$, let $\ket{\Phi}_{RA}$ be maximally entangled with a subspace of $A$ and set $\ket{\psi}_{RBB_c} = (\openone_R \otimes V) \ket{\Phi}_{RA}$.
Then there is a ``decoding'' channel $\B$ from $B$ to $R' \cong R$ satisfying
\begin{equation}
    \left\| \Phi_{RR'} - \big(\id_R\otimes\B\circ\N\big)(\Phi_{RA}) \right\|_1
    \leq 2 \left\|\psi_{RB_c} - \Phi_R\otimes \psi_{B_c}\right\|_1^{1/2}.
\end{equation}
\end{lem}
\begin{proof}
This is a widely used fact in quantum Shannon theory.  The proof is
similar to that of Lemma \ref{lem:duality.decoupling}.  For details,
see Theorem II of \cite{Hayden08}, which is an equivalent statement
up to an application of Eq.~(\ref{eqn:trace.fid}).

\end{proof}

\begin{proof} \emph{Converse part of Theorem \ref{thm:qpriv}.}
To prove optimality, suppose we have an $(n, \lfloor nQ
\rfloor,\delta,\epsilon)$ private entanglement transmission code.  As
before, use $R$ to denote the reference space for the maximally
entangled state $\Phi_{2^k}$ in the definition, with $k = \lfloor nQ
\rfloor$.  Let $\ket{\Psi}_{RFE^n\Ecb^n}$ be the purified final state
after $\E^{\otimes n} \circ \A$ has acted on $\Phi_{2^k}$.  The
privacy condition  $\| \Psi_{RE^n} - \Psi_{R} \otimes \Psi_{E^n}
\|_1 \leq \epsilon$ and Lemma \ref{lem:privacy.entanglement} imply
that there exists a ``decoding''  channel $\D$ on $\Ecb^n F$ such
that
\begin{equation}
    \left\| \Phi_{2^k} - (\id_{R} \otimes \D)(\Psi_{RFE_c^n}) \right\|_1 \leq 2 \sqrt{\epsilon}.
\end{equation}

The Alicki-Fannes inequality (Lemma \ref{lem:af.inequality}) then
implies that there is a function $g_1(\epsilon)$ satisfying
$\lim_{\epsilon \rightarrow 0} g_1(\epsilon) = 0$ such that
\begin{equation} \label{eqn:converse.1}
    2 \lfloor nQ \rfloor
     = I(R;A)_{\Phi_{2^k}}
    \leq I(R;A)_{(\id_{R} \otimes \D)(\Psi)} + ng_1(\epsilon).
\end{equation}
The monotonicity of the mutual information under quantum channels
then implies that
\begin{eqnarray}
    I(R;A)_{(\id_{R} \otimes \D)(\Psi)}
    &\leq& I(R;\Ecb^n F)_\Psi \\
    &=& I(R;F)_\Psi + I(R;\Ecb^n|F)_\Psi,  \label{eqn:converse.2}
\end{eqnarray}
where the second line is just the chain rule for mutual information.
Now consider $I(R;F)_\Psi$.  The entanglement transmission condition
requires that
\begin{equation}
 \left\|
  (\id \otimes \B \circ \N^{\otimes n} \circ \A)
    ( \Phi_{2^k} ) -
     \Phi_{2^k}
 \right\|_1 \leq \delta
\end{equation}
Let $\ket{\Omega}_{RFB_c^nC{\Cc}}$ be a purification of $(\id_{R}
\circ \B \circ \N^{\otimes n} \circ A)(\Phi_{2^k})$, where ${\Bc}$
is the environment of $\N^{\otimes n}$ and ${\Cc}$ the environment
of $\B$.  The entanglement transmission condition and Lemma
\ref{lem:duality.decoupling} together imply that there is a state
$\xi_{F{\Bcb}^n{\Cc}}$ such that
\begin{eqnarray}
    2\sqrt{\delta}
    &\geq& \left\| \Omega_{RF{\Bcb}^nC{\Cc}} - (\Phi_{2^k})_{RC} \otimes \xi_{F{\Bcb}^n{\Cc}} \right\|_1 \\
    &\geq& \left\| \Omega_{RF} - \pi_{R} \otimes \xi_{F} \right\|_1,
\end{eqnarray}
where the second inequality is a consequence of the monotonicity of
the trace distance under the partial trace.  But $\Psi_{RF} =
\Omega_{RF}$ since neither $\E^{\otimes n}$ nor $\B \circ
\N^{\otimes n}$ acts on $RF$.  So, again by the Alicki-Fannes
inequality,  there is a function $g_2(\delta)$ satisfying
$\lim_{\delta \rightarrow 0} g_2(\delta) = 0$ such that
\begin{equation} \label{eqn:converse.3}
    I(R;F)_\Psi \leq n g_2(\delta).
\end{equation}
Combining Eqs.~(\ref{eqn:converse.1}), (\ref{eqn:converse.2}) and
(\ref{eqn:converse.3}) then gives
\begin{equation}
    2 \lfloor nQ \rfloor \leq I(R;\Ecb^n|F)_\Psi + n[g_1(\epsilon) + g_2(\delta) ].
\end{equation}
But
\begin{equation}
    I(R;\Ecb^n|F)_\Psi = I(RF;\Ecb^n)_\Psi - I(\Ecb^n;F)_\Psi \leq I(RF;\Ecb^n)_\Psi
\end{equation}
by the chain rule and the nonnegativity of mutual information.  Thus,
we finally arrive at the conclusion that
\begin{equation}
    2 \lfloor nQ \rfloor \leq I(RF;\Ecb^n)_\Psi + n[g_1(\epsilon) + g_2(\delta) ].
\end{equation}
The composite system $RF$ can be thought of as the purification of
the input to the channel, which is the role played by $A'$ in
Theorem \ref{thm:qpriv}.  Relabeling $RF$ by $A'$ and recalling that
the inequality must hold for all $\delta, \epsilon > 0$ and $n$
sufficiently large then shows that
\begin{equation}
Q_p(\id, \E) \leq \lim_{n \rightarrow \infty} \max \frac{1}{2n}
I(A';\Ecb^n)_\rho,
\end{equation}
where the maximization is over pure states $\ket{\psi}_{A'^nA^n}$
and $\rho = (\id \otimes \E_c^{\otimes n})(\psi)$.  It is
well-known, however, that fixing $n=1$ does not affect the
expression on the right hand side of the inequality, which is the
entanglement-assisted quantum capacity of $\E_c$~\cite{BSST02}. That
completes  the proof of the converse.
\end{proof}

\section{The qudit Unruh channel: Definition and structure}
In this section we define the qudit Unruh channel and determine the
structure of the output density matrix.  One of the key consequences of the
structure theorem will be the covariance of the qudit Unruh channel with
respect to the $SU(d)$ group.

\subsection{The Unruh effect}
\label{subsec:Unruh}
In order to describe the Unruh effect it will be useful to briefly
recapitulate the construction of a quantum field theory.  One begins
with the classical field theory and its space of solutions.  One
uses the ``time'' coordinate to define a space of positive-frequency
solutions, this is taken as the Hilbert space of ``one-particle''
states, $\H$.  One then constructs the usual Fock space, $\F(\H)$
over this Hilbert space.  This Fock space comes with its usual
apparatus of annihilation and creation operators, $a_k,a_k^\dg$
respectively.  The vacuum state is the unique state killed by all
the $a_k$.

\begin{figure}[t]
\begin{center}
    \resizebox{7cm}{6cm}{\includegraphics{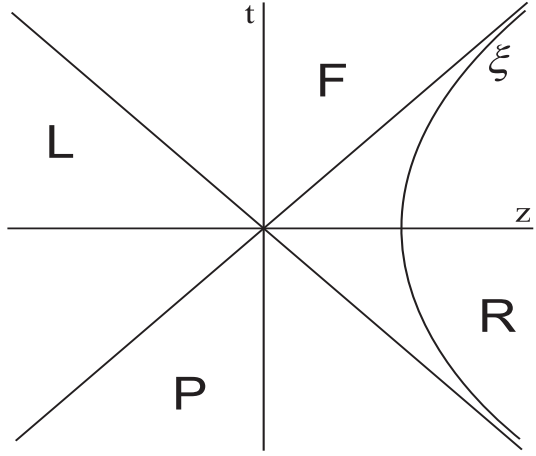}}
    \caption{Illustration of a timelike Killing field $\xi$ chosen for a uniformly accelerating observer. The letters F, R, P and L stand for future, right, past and left cone, respectively.}
    \label{fig:sud_plainspacetime}
\end{center}
\end{figure}

In Minkowski space one has the usual quantization procedure based on
the usual timelike Killing field that yields a Hilbert space $\H_M$
with a Fock space $\F(\H_M)$ and a vacuum state that we call
$\ket{vac}_M$.  We can, however, use another timelike Killing field
$\xi$, the one whose integral curves are the trajectories of an
accelerating observer.  For such an observer the spacetime consists
of 4 regions as shown in Figure~\ref{fig:sud_plainspacetime}.  If we look
at positive frequency states with respect to this notion of time, we
can divide them into solutions that live in the left wedge and those
that live in the right wedge.  We get two Hilbert spaces,
$\H_L,\H_R$ and their respective Fock spaces $\F(\H_L),\F(\H_R)$.
The space of one-particle states appropriate to the accelerating
observer, we shall call her the Rindler observer, is $\H_{Rin}
\eqsim\H_L\oplus\H_R$ We have that $\F(\H_{Rin})\eqsim
\F(\H_L)\otimes\F(\H_R)$.  The transformation from the Minkowski
observer's Fock space to the Rindler observer's Fock space is given
by a map $S:\F(\H_M)\to\F(\H_{Rin})$.  The Minkowski vacuum
$\ket{vac}_M$ will appear as $S\ket{vac}_M$ in the quantum field
theory of the Rindler spacetime The accelerating observer can only
perceive states of $\F(\H_R)$ so the correct description of how she
perceives the state is obtained by tracing out the states of
$\F(\H_L)$.  Thus, she sees a mixed state.
The transformation $S$ takes the vacuum state $\ket{vac}_M$ to an
infinite product state corresponding to all possible modes, and in
fact the Minkowski and Rindler field theories are not unitarily
equivalent~\cite{WaldBook}.  (This may appear to contradict the statement
above that there is a map $S$ between the Fock spaces.  The point is that
there is no such \emph{unitary} $S$ between the indicated Fock spaces.)


In this paper we will use a fixed number of input modes and restrict
attention to a fixed number of modes of the output rather than all possible
modes.  Physically we can think of the Rindler observer's detector being
tuned to some \emph{finite number of modes}.  The Fock space that we get
this way is unitarily equivalent to the input Fock space also restricted to
these modes, so we can define a unitary transformation between the input
and output spaces, where the output is the restricted Fock space for both
Rindler wedges.  We now proceed with the mathematical details.

The solution of the Klein-Gordon equation for a real massless scalar field
in Minkowski spacetime can be expanded in terms of the so-called Unruh
modes $U_{\pm\Om,\bk}$~\cite{crispy}
\begin{equation}\label{eq:unruhmode_function}
\phi_{Unr}=\int_{0}^\infty d\Om\int_{-\infty}^\infty d\bk
\left[
d_{\Om,\bk}U_{\Om,\bk}+d^{\dg}_{\Om,\bk}\bar U_{\Om,\bk}
+d_{-\Om,\bk}U_{-\Om,\bk}+d^{\dg}_{-\Om,\bk}\bar U_{-\Om,\bk}
\right].
\end{equation}
The bar denotes complex conjugation and the field coefficients
$d_{\pm\Om,\bk},d^{\dg}_{\pm\Om,\bk}$ are the (Unruh) bosonic creation
and annihilation operators satisfying
$\big[d_{\pm\Om,\bk},d^{\dg}_{\pm\Om',\bk'}\big]=\d(\bk-\bk')\d(\Om-\Om')$
with any other combination equal to zero. Similarly, if we introduce
the spacetime of a uniformly accelerating observer (Rindler spacetime) the
same field can be expanded in terms of the left and right Rindler modes
$R_{\Om,\bk}^\pm$
\begin{equation}\label{eq:rindlermode_function}
\phi_{Rin}=\int_0^\infty d\Om\int_{-\infty}^\infty d\,\bk\left[b^R_{\Om,\bk}R_{\Om,\bk}^++b^{R^\dg}_{\Om,\bk}\bar R_{\Om,\bk}^++b^L_{\Om,\bk}R_{\Om,\bk}^-+b^{L^\dg}_{\Om,\bk}\bar R_{\Om,\bk}^-\right].
\end{equation}
The number $\Om$ is the mode frequency divided by Rindler observer's
proper acceleration and $\bk$ is the mode three-momentum.

The relation between the different modes is as
follows~\cite{unruh76,crispy}.  One can define Minkowski modes using plane waves
according to the standard Minkowski coordinates.  The Unruh modes
of Eq.~(\ref{eq:unruhmode_function}) are linear combinations of Minkowski
modes. They are, however, parametrized by the quantities $\Om$ and $\bk$
that refer to the accelerated observer.  The annihilation operators
associated with the Unruh modes are linear combinations of annihilation
operators of the Minkowski modes, i.e. they are not mixed with the
Minkowski creation operators and they also annihilate the Minkowski vacuum.
They serve as a convenient intermediate set of modes in going from the
Minkowski field theory to the Rindler field theory.  The Rindler modes are
the ones defined using the positive and negative frequency decomposition of
the accelerating observer.  When related to the Minkowski modes they will
have a mixture of positive and negative frequency Minkowski modes.  The
equations above give the expansion of the \emph{same} field operator in
terms of the two different modes.


The Rindler annihilation and creation operators come in two pairs
associated with the left and right wedges; the ones associated with the
right wedge are denoted by $b^R_{\Om,\bk}$ and $b^{R^\dg}_{\Om,\bk}$. They
separately satisfy the same commutation relations as the Unruh
operators.  Comparing both expressions for what are in fact the same the field operators we get the
Bogoliubov transformation between the Unruh and Rindler creation and
annihilation operators
\begin{equation}\label{eq:symplec}
\begin{pmatrix}
  b^R_{\Om,\bk} \\
  {b^L}^\dg_{\Om,-\bk} \\
\end{pmatrix}
=\begin{pmatrix}
   \cosh{r} & \sinh{r} \\
   \sinh{r} & \cosh{r} \\
 \end{pmatrix}
 \begin{pmatrix}
  d_{-\Om,\bk} \\
  {d^\dg}_{\Om,-\bk} \\
 \end{pmatrix},
\end{equation}
where
$\cosh{r}=\sqrt{e^{\Om}/(e^{\Om}-1)},\sinh{r}=\sqrt{1/(e^{\Om}-1)}$~\cite{crispy,WaldBook}.
We use the natural units $\hbar=c=1$. The transformation completely
describes the physics of a uniformly accelerated observer. We are
able to calculate the expectation values of any Rindler operator in
terms of the Unruh modes.  The celebrated thermal spectrum of the
Minkowski vacuum as seen by the Rindler observer is an example of
such a calculation. In Eq.~(\ref{eq:symplec}) we witness another
advantage of working with Unruh modes: the transformation between
Unruh and Rindler modes is very simple.

Inverting Eq.~(\ref{eq:symplec}) we can see that every Unruh Fock state
can be expanded as a function of the left and right Rindler modes.  In other
words, there is an operation $\O$ assigning a two-mode entangled Rindler
state to every state from Minkowski spacetime:
\begin{equation}\label{eq:MinkRinassignement}
    \O:\ket{n}_{Unr}\mapsto\prod_{\Om,\bk}{1\over\cosh^{1+n}{r}}
    \sum_{m=0}^\infty\binom{n+m}{n}^{1/2}\tanh^m{r}\ket{(n+m)^L_{\Om,-\bk}}_{Rin}\ket{m^R_{\Om,\bk}}_{Rin}.
\end{equation}
The mathematical reason that the transformation between the two Fock
representations is not unitary~\cite{WaldBook} is the presence of the
infinite product.  An obvious way to circumvent this problem is to restrict
to just one output mode of the operation $\O$.  We would like to find a
unitary operation ``emulating'' the action of the restricted $\O$.
Effectively, it is the same as introducing a two-mode unitary
transformation
\begin{equation}\label{eq:UnruhUnitary}
    U_{AC}(r)=\exp\big[r( a^\dg c^\dg - ac)\big].
\end{equation}
This formulation deliberately uses a different mode notation (the
labels $A\equiv(\Om,\bk)$ and $C\equiv(\Om,-\bk)$ ), the reason
being that the output mode restriction allows us to work in a single
Hilbert space. We stress that the unitary transformation
Eq.~(\ref{eq:UnruhUnitary}) produces the ``correct''
states~Eq.~(\ref{eq:MinkRinassignement}) as seen by a Rindler
observer. The shortened notation also avoids carrying too many
indices with all the mode information. The two different symbols $A$
and $C$ correspond to the operators $a$ and $c$, respectively.
Therefore $U_{AC}(r)$ acts as
\begin{equation}\label{eq:UnruhUnitary_eff}
    U_{AC}(r)\ket{n}_A\ket{vac}_C={1\over\cosh^{1+n}{r}}\sum_{m=0}^\infty\binom{n+m}{n}^{1/2}\tanh^m{r}\ket{n+m}_{A}\ket{m}_{C}.
\end{equation}

Now suppose we want to transform an arbitrary (pure) qudit. There
are many ways  to encode a logical qudit, but we will restrict to a natural method
known as multi-rail  encoding, in which
an arbitrary qudit state takes the form
\begin{equation}\label{eq:multi-railbasis}
    \ket{\psi}_{A}=\sum_{i=1}^d\beta_ia_i^\dg\ket{vac}_A.
\end{equation}
In other words, there are $d$ distinguishable modes and the unitary acts on
each mode to give
\begin{equation}\label{eq:unitaryaction}
\ket{\s}_{AC}=\bigotimes_{i=1}^dU_{A_iC_i}\ket{\psi}_{AC},
\end{equation}
where $\ket{\psi}_{AC}=\ket{\psi}_{A}\ket{vac}_C$. The disentangling theorem allows us to rewrite the exponential as~\cite{barnett1997methods}
\begin{multline}\label{disentang}
    U_{A_iC_i}(r)={1\over\cosh{r}}\exp{\big[\tanh{r}\,a_i^\dg c_i^\dg\big]}\\
    \times\exp{\big[-\ln{\cosh{r}}(a_i^\dg a_i+c_i^\dg c_i)\big]}\exp{\big[-\tanh{r}\,a_ic_i\big]}.
\end{multline}
Using the commutation relations
$[a_i^\dg,a_j^\dg]=[a_i,c_j^\dg]=[a_i,c_j]=0$, the action of
$U_{AC}$ simplifies to
\begin{equation}\label{eq:unitary_product_simple}
    U_{AC}\ket{\psi}_{AC}=\bigotimes_{i=1}^dU_{A_iC_i}\ket{\psi}_{AC}={1\over\cosh^{d+1}{r}}\exp{\left[\tanh{r}\left(\sum_{i=1}^da_i^\dg
          c_i^\dg\right)\right]}\ket{\psi}_{AC}.
\end{equation}
Note that this simplification holds only when $U$ transforms states
from the Hilbert space spanned by the multi-rail basis.  The
summands in the Taylor series for $U$ can be simplified by the
multinomial theorem to give
\begin{equation}\label{Taylor_summands}
    {\tanh^k{r}\over k!}\left(\sum_{i=1}^da_i^\dg c_i^\dg\right)^k
    =\tanh^k{r}\sum_{\ahead{l_1+\hdots+l_d=k}{}}{1\over l_1!\dots
      l_d!}(a_1^\dg c_1^\dg)^{l_1}\dots(a_d^\dg c_d^\dg)^{l_d}.
\end{equation}
The simplified expression Eq.~(\ref{eq:unitary_product_simple}) allows us to
rewrite Eq.~(\ref{eq:unitaryaction}) in the following way:
\begin{multline}\label{transformed_qudit}
    \ket{\s}_{AC}=\left(\sum_{i=1}^d\beta_ia_i^\dagger\right)U\ket{vac}\\
    ={1\over\cosh^{d+1}{r}}\left(\sum_{i=1}^d\beta_ia_i^\dagger\right)\sum_{k=0}^\infty\tanh^k{r}\sum_{\ahead{l_1+\hdots+l_d=k}{}}
    \ket{l_1\dots l_d}_A\ket{l_1\dots l_d}_C,
\end{multline}
where $1/\sqrt{l_i!}(a_i^\dg)^{l_i}\ket{vac}=\ket{l_i}$ has been
used in the second line. Thus, an input state $\ket{\psi}_A$ of the
from Eq.~(\ref{eq:multi-railbasis}) gets transformed to a final
output state
\begin{equation}\label{eq:transformed_qudit_final}
    \ket{\s}_{AC}={1\over\cosh^{d+1}{r}}\sum_{k=1}^\infty\tanh^{k-1}{r}
    \sum_{I}\left[\sum_{i=1}^d\beta_i\sqrt{l_{I,i}+1}\ket{I^{(i)}}_A\ket{I}_C\right],
\end{equation}
where $\ket{I}_C=\ket{l_1\dots l_d}_C$ is a multi-index labeling for basis
states of the completely symmetric subspace of $(k-1)$ photons in $d$
modes. Note that $k$ was relabeled as $k+1$ so in comparison with
Eq.~(\ref{transformed_qudit}) we now have $k=\sum_{i=1}^d l_{I,i}+1$. A ket
$\ket{I^{(i)}}_A$ differs from $\ket{I}_C$ by having $l_{I,i}+1$ instead of
$l_{I,i}$ in the $i$-th place, that is, $\ket{I^{(i)}}_A=\ket{l_{I,1}\dots
  l_{I,i}+1\dots l_{I,d}}_A$.  The interpretation is that the $A$ subsystem contains
$k$ photons in $d$ modes.  The presence of the index $I$ is crucial since
the value of $l_{I,i}$ indeed depends on which $\ket{I}_C$ was used to
generate the corresponding $\ket{I^{(i)}}_A$.
\begin{exa}
    For $d=3$ and $k=2$ the basis consists of the states \\
    $\big\{\ket{I}_C\big\}=\big\{\ket{001},\ket{010},\ket{100}\big\}$
    corresponding to a single photon in three possible modes of the
    $A$ subsystem. For $\ket{100}_C$ we get
    $\big\{\ket{I^{(i)}}_A\big\}_{i=1}^3=\big\{\ket{200},\ket{110},\ket{101}\big\}$
    with the coefficient $l_{I,i}+1$ equal to $2,1$ and $1$,
    respectively. If we chose a different $\ket{I}_C$ the result would in
    general be a different set of vectors and coefficients.
\end{exa}
\begin{exa}
    For another example we choose $d=4$ and $k=3$. The basis of the $C$~subsystem consists of the states
\begin{multline*}
\big\{\ket{I}_C\big\}=\big\{\ket{0002},\ket{0020},\ket{0200},\ket{2000},\ket{0011},\ket{0101},\ket{0110},\\
\ket{1001},\ket{1010},\ket{1100}\big\}
\end{multline*}
This corresponds to two photons in four possible modes of the $A$~subsystem. For $\ket{0200}_C$ we get
$\big\{\ket{I^{(i)}}_A\big\}_{i=1}^4=\big\{\ket{1200},\ket{0300},\ket{0210},\ket{0201}\big\}$
with the coefficient $l_{I,i}+1$ equal to $1,3,1$ and $1$,
respectively.
\end{exa}

We conclude this section by providing a quantum-optical interpretation of
Eq.~(\ref{eq:UnruhUnitary_eff}). In this setting the situation is
conceptually simpler than the Unruh effect since the issue with
non-equivalent Hilbert spaces does not arise. This is indicated by an
isometric identification of the input and output Hilbert space in
Eq.~(\ref{eq:UnruhUnitary_eff}) (the identical labels on the LHS and
RHS). Consider an array of $d$ two-mode optical squeezers each described by
the unitary operation of Eq.~(\ref{eq:UnruhUnitary}). The overall action is
described by Eq.~(\ref{eq:unitaryaction}) since all the input $C$-modes
(the environment) contain vacuum. The total number of photons in the input
$A$-mode equals one and the input pure state can be described by an
arbitrary superposition of $d$ modes as in
Eq.~(\ref{eq:multi-railbasis}). The squeezing parameter $r$ is common for
all squeezing transformations and is analogous to the proper acceleration
of the Rindler observer. Simpler constructions of this kind were previously
investigated in connection with quantum optical
amplifiers~\cite{braunstein2001optimal,caves}.

\subsection{The structure of the output density matrix and the irreducible representations of $\slnc$}
\label{subsec:geomstruct}

We will show that the terms appearing in the output density matrix
live in spaces that carry representations of $\slnc$ and the states
themselves can be written in terms of the Lie algebra elements.

We begin with a formal definition of the qudit Unruh channel.
\begin{defi}\label{def:Unruhchannel}
    The qudit Unruh channel $\E$ is the quantum channel defined by
    $\E(\psi_{A})= \Tr{C} U \psi_{AC} U^\dagger$  where
    $U=\bigotimes_{i=1}^dU_{A_iC_i}$ with $U_{A_iC_i}$ given by
    Eq.~(\ref{eq:unitaryaction}). The action of the channel on an input
    qudit state~Eq.~(\ref{eq:multi-railbasis}) is given by
    \begin{equation}\label{eq:Unruhchanneloutput}
        \E:\psi_A\mapsto\sigma_A=(1-z)^{d+1}\bigoplus_{k=1}^\infty
        z^{k-1}\sigma_A^{(k)},
    \end{equation}
    where
    \begin{equation}\label{complete_state_traced}
        \begin{split}
        \sigma_A^{(k)} = & \ \sum_{I}
        \sum_{i=1}^d|\b_i|^2(l_{I,i}+1)\kb{I^{(i)}}{I^{(i)}}_A \\
        + & \ \sum_{I}
        \sum_{\ahead{i,j=1}{i\not=j}}^d\b_i\bar\b_j\sqrt{(l_{I,i}+1)(l_{I,j}+1)}
        \kb{I^{(i)}}{I^{(j)}}_A,
        \end{split}
    \end{equation}
    and we have defined $z=\tanh^2{r}$ so that ${\cosh^2{r}}=1/(1-z)$.
\end{defi}
\begin{rem}
  From Eq.~(\ref{complete_state_traced}) we find
  that ${\sigma_A^{(k)}}$ has ${d+k-1\choose k}$ rows and columns, while
  $\Tr{}\sigma_A^{(k)}={d+k-1\choose d}$, which leads to $\Tr{} \sigma_A = 1$.
  Note also that the letters $A$ and $C$ are
  used for labelling both the input and output systems.
\end{rem}

In summary, the qudit Unruh channel is a map transforming states
prepared in a limited sector of the Minkowski observer's Hilbert
space (the observer we have called Alice) to the Hilbert space
associated with a uniformly accelerating observer (Rindler observer
Eve).

\begin{thm}\label{thm:invariant_structure}
Let $\lambda_{\alpha}$ be the generators of the $\slnc$ Lie algebra in the
Chevalley-Serre basis (defined in Appendix A).  We write
$\lambda_{\alpha}^{(k)}$ for the matrix representation of
$\lambda_{\alpha}$ in the $k$th completely symmetric representation.
There exist numbers
$n_{\alpha}$, \emph{that are independent of} $k$, such that each block of
the output density matrix $\sigma_A$ can be written in the form
    \begin{equation}\label{eq:allblocks}
    \s_A^{(k)}={1\over d}\Big(k\openone+\sum_{\a=1}^{L}n_\a\lam^{(k)}_\a\Big).
    \end{equation}
In particular the first block of $\s_A$ in
Eq.~(\ref{eq:Unruhchanneloutput}) can be written as
    \begin{equation}\label{eq:firstblock}
    \s_A^{(1)}={1\over d}\Big(\openone+\sum_{\a=1}^{L}n_\a\lam^{(1)}_\a\Big),
    \end{equation}
    where $\lam^{(1)}_\a$ are generators of the fundamental representation
    of the $\slnc$ algebra defined in the appendix and $L=2d(d-1)$.

\end{thm}
\begin{rem} The set of
generators is overcomplete since, for all
$d\geq2$,  one has $L=2d(d-1)>d^2-1 = \dim \slnc$
 so there is not an unique expansion of $\sigma_A$; however, the
theorem asserts that there is \emph{a particular choice} of the expansion
coefficients such that the same coefficients can be used for all the
blocks; these will be described explicitly in the proof.
The use of this overcomplete set of  $\slnc$ algebra generators will make
it easier to identify the components of Eq.~(\ref{eq:allblocks}) with the $\slnc$-algebra generators for all $k$.
It is more convenient than working with a basis but not essential.
\end{rem}

\begin{rem}
  The blocks $\s_A^{(k)}$ are generally not normalized.
  The factor multiplying the identity
  in Eq.~({\ref{eq:allblocks}}) comes from the ratio
  $\Tr{}\openone/\Tr{}\s_A^{(k)}={d+k-1\choose k}/{d+k-1\choose d}=d/k$.
\end{rem}

The proof of Theorem~\ref{thm:invariant_structure}
will be split into two lemmas.  Lemma~\ref{lem:diagk1} will handle the
diagonal coefficients of Eq.~(\ref{eq:allblocks}) and
Lemma~\ref{lem:offdiagk1} will address the off-diagonal coefficients.
The formalism that we use is called the boson operator formalism or boson
calculus.

This is a convenient algebraic formalism that relates the matrices that
arise from the representations of $\slnc{}$ with the boson annihilation and
creation operators.  Since bosons are invariant under the action of
permutations, the boson algebra is well adapted to describing the
completely symmetric representations that we care about.  The presentation
here follows the exposition in the text by Gilmore~\cite{gilmore}.

If one is working with $n\times n$ matrices then one introduces $n$
independent commuting pairs of boson creation $a_i^{\dagger}$ and
annihilation $a_i$ operators.  These obey the commutation rule
\[ [a_i,a_j^{\dagger}] = \delta_{ij} I\] where $I$ is the identity operator
of the algebra generated by the boson operators.  Given a matrix $A$ we
define a corresponding bilinear operator $\hat{A}$ by the rule
\[ \hat{A} = \sum_{ij}a_i^{\dagger}A_{ij}a_j.  \]
An elementary calculation shows that
\[ [\hat{A},\hat{B}] = \widehat{[A,B]}\]
where the bracket on the left is the commutator in the boson algebra and
the one on the right is the ordinary commutator of matrices.

This correspondence shows that the commutators of the matrices are
faithfully represented by the commutators of the operators; note, however,
that the correspondence does not respect ordinary multiplication.  Thus,
given the commutation relations of a Lie algebra, we can hope to represent
them using appropriate combinations of boson operators.  In fact, such a
correspondence works for all the $\slnc{}$ algebras and for some other
classes of Lie algebras as well.  This was originally discovered by Jordan
and rediscovered several years later by Schwinger in the case of $\sltc$,
where it is called the Schwinger oscillator representation of angular
momentum.

With these notations in place we have the following theorem whose proof is
a routine calculation.
\begin{thm}[See \cite{gilmore}]\label{thm:bosrep}
Let $a_1,a_1^{\dg},\ldots,a_d,a_d^{\dg}$ be $d$ pairs of operators obeying
the canonical commutation relations for bosonic annihilation and creation
operators.  Then for $1\leq l,m\leq d$, the following operators
    \begin{subequations}\label{eq:bosrep_of_sl2C}
        \begin{align}
            \widehat{H_{ml}} &= a_m^\dg a_m-a_{l}^\dg a_{l} \\
            \widehat{E_{ml}} &= a_m^\dg a_l\label{eq:bosrep_of_sl2Cb}\\
            \widehat{E_{ml}^\dg} &= a_l^\dg a_m,
        \end{align}
    \end{subequations}
    obey the commutation relations of the $\slnc$ algebra.  That is, the
    operators satisfy the commutation relations (\ref{eq:sl2cinChevalley})
    from the appendix.
\end{thm}
The operator
\begin{equation}
  \widehat{\openone}  \stackrel{\text{def}}{=}\sum_{i=1}^da^\dg_ia_i,
\end{equation}
commutes with all the operators from Eqs.~(\ref{eq:bosrep_of_sl2C}); it is
the operator corresponding to an identity matrix and it is also the operator representing an identity from~Eq.~(\ref{eq:firstblock}) for all $d$.

The above correspondence is well adapted to dealing with the fundamental representation.  This
corresponds to the case where there is one photon in one of $d$ modes: in
the boson operator form we see that there is exactly one pair of boson
operators for each mode.  The completely symmetric representations
corresponding to higher $k$ require a mild generalization of the
correspondence between matrices and boson operators where we have
higher-index tensors instead of matrices.

Let $a_1,a_1^{\dg},\ldots,a_d,a_d^{\dg}$ be $d$ pairs of operators obeying
the canonical commutation relations for bosonic annihilation and creation
operators.  Suppose that $V$ is the fundamental representation of $\slnc$;
hence it is a vector space carrying a $d$-dimensional representation of
$\slnc$.  Let $e_1,\ldots,e_d$ be the basis vectors in some basis for $V$.
Then the basis vectors of the representation formed by the completely
symmetrized $k$-fold tensor product of $V$ will be labeled by indices
$1\leq i_1\ldots,i_k\leq d$.  The action of the $\slnc$ Lie algebra
elements will be given by tensors with $2k$ indices $1\leq
i_1\ldots,i_k,j_1,\ldots,j_k\leq d$.  The first $k$ indices are completely
symmetrized and the next $k$ are also completely symmetrized.  Then we make
the following correspondence between operators and tensors $A$ (acting on
the completely symmetric representation)
\begin{equation}\label{eq:bosrepk}
A_{i_1,\ldots,i_k,j_1,\ldots,j_k}\mapsto
\sum_{i_1,\ldots,i_k,j_1,\ldots,j_k} a_{i_1}^{\dg}\ldots a^{\dg}_{i_k}
A_{i_1,\ldots,i_k,j_1,\ldots,j_k} a_{j_1}\ldots a_{j_k}.
\end{equation}

This notation makes it easy to write a bosonic representation of an identity from  Eq.~(\ref{eq:allblocks}).
The corresponding tensor is
$
    A_{i_1,\ldots,i_k,j_1,\ldots,j_k} = \d_{i_1j_1}\dots\d_{i_kj_k}
$
and it takes the form of a ${d+k-1\choose k}$-dimensional matrix with ones on the diagonal and zeros everywhere else.

To help the reader more easily follow the upcoming lemmas, we
provide some examples which illustrate how
Eq.~(\ref{eq:transformed_qudit_final}) leads to the block diagonal
structure of the output density matrix of
Eq.~(\ref{complete_state_traced}).
\begin{exa}
For $k=1$ and an arbitrary $d$ we get from
Eq.~(\ref{eq:transformed_qudit_final})
\begin{align}\label{eq:example_dk1}
    \sum_{I}\Bigg[\sum_{i=1}^d\beta_i&\sqrt{l_{I,i}+1}\ket{I^{(i)}}_A
\ket{I}_C\Bigg]\nn\\
    &=\left(\b_1\ket{1\dots0}_A+\b_2\ket{01\dots0}_A+\dots+\b_d\ket{0\dots1}_A
\right)\ket{vac}_C.
\end{align}
\end{exa}
\begin{exa}
Let $d=3$ and $k=3$. Picking up the relevant part of
Eq.~(\ref{eq:transformed_qudit_final}) gives
\begin{align}\label{eq:exampled3k3}
    \sum_{I}\Bigg[\sum_{i=1}^d\beta_i\sqrt{l_{I,i}+1}&\ket{I^{(i)}}_A\ket{I}_C\Bigg]\nn\\
    &=\Big[\left(\b_1\ket{102}_A+\b_2\ket{012}_A+\b_3\sqrt{3}\ket{003}_A\right)\ket{002}_C\nn\\
    &+\left(\b_1\ket{120}_A+\b_2\sqrt{3}\ket{030}_A+\b_3\ket{021}_A\right)\ket{020}_C\nn\\
    &+\left(\b_1\sqrt{3}\ket{300}_A+\b_2\ket{210}_A+\b_3\ket{201}_A\right)\ket{200}_C\nn\\
    &+\left(\b_1\ket{111}_A+\b_2\sqrt{2}\ket{021}_A+\b_3\sqrt{2}\ket{012}_A\right)\ket{011}_C\nn\\
    &+\left(\b_1\sqrt{2}\ket{201}_A+\b_2\ket{111}_A+\b_3\sqrt{2}\ket{102}_A\right)\ket{101}_C\nn\\
    &+\left(\b_1\sqrt{2}\ket{210}_A+\b_2\sqrt{2}\ket{120}_A+\b_3\ket{111}_A\right)\ket{110}_C\Big].
\end{align}
\end{exa}
\begin{lem}\label{lem:diagk1}
    The diagonal part of Eq.~(\ref{eq:allblocks}) can be written as a sum of
    the diagonal generators of the $k$-th completely symmetric
    representation of $\slnc$ algebra as specified by Theorem~\ref{thm:invariant_structure}, and for a given $d$ the coefficients
    $n_\a$ of the diagonal algebra generators can be chosen independently
    of the representation.
\end{lem}
\begin{proof}
    \noindent\textbf{Case} $k=1$.
    Eq.~(\ref{eq:example_dk1}) from the first example makes it clear
    that the diagonal part of $\sigma_A^{(1)}$ can be written as
    \begin{equation}
        \sum_{i=1}^d |\beta_i|^2 \ket{0\ldots 1_i\ldots
          0}\bra{0\ldots 1_i\ldots 0},
    \end{equation}
    where the $1$ appears in the $i$th position.  Now the matrix
    $\proj{0\ldots 1_i\ldots 0}$ under the boson correspondence gives the
    operator $\boson{i}$.  It therefore suffices to rewrite $a_i^\dagger a_i$
    using the boson algebra and then transform back to matrices.  We will do
    the case where $i=1$ since other values of $i$ are identical up to cyclic
    permutations of the indices:
    \begin{equation}\label{eq:beta1op}
        a_1^\dg a_1
            = \frac{1}{d} \Big(\sum_{k=1}^da^\dg_ka_k+\sum_{j=1}^{d-1}(a_1^\dg a_1-a_{1+j}^\dg a_{1+j})\Big)
            = \frac{1}{d} \left( \widehat{\openone} + \sum_{j=1}^{d-1} \widehat{H_{1,1+j}}
            \right).
    \end{equation}

    Now we get back to the matrix form by ``removing the hats'', i.e.\ we have
    \begin{equation}\label{eq:beta1}
    \proj{1\ldots 0} = \frac{1}{d} \left( \openone +
      \sum_{j=1}^{d-1} H_{1,1+j}\right).
    \end{equation}
    Taking into account the $\beta$ coefficients we see that $d$ of the
    $n_{\alpha}$ given by tracing over the $C$ subsystem in~Eq.~(\ref{eq:example_dk1}) are just $|\beta_i|^2$ for $i=1,\ldots,d$.
    \begin{rem}
      The diagonal generators of the form $H_{i,i+1}$, with
      $\widehat{H_{i,i+1}} = a_i^\dg a_i-a_{i+1}^\dg a_{i+1}$, correspond to
      the simple roots of the fundamental representation introduced in the
      appendix while the rest are their linear combinations.  As mentioned
      earlier, because of the use of a linearly dependent set of diagonal
      generators, the expansion coefficients are not uniquely determined.
      However, for the explicit choice of expansion coefficients and generators
      given here, the coefficients are independent of the representation.
    \end{rem}

    \noindent\textbf{Case} $k>1$.

It can be readily observed that squaring the coefficient accompanying
$\b_i$ gives the label of the $i$-th ket of the $A$ subsystem.  So whatever
combination leads to the diagonal part corresponding to $\b_1$ will lead
after a suitable simple modification to the diagonal part corresponding to
$\b_i$.

Using the form of the diagonal generators of $\slnc$ in the Chevalley-Serre
basis (presented in the appendix) we may conclude certain general things
about the form of the representations of the diagonal operators in the
completely symmetric representations for all $k$ in
Eq.~(\ref{complete_state_traced}).  Recall that in the geometric picture of the
root space each line of the root space diagram is some embedded
representation of the $\sltc$ algebra. Following the construction of
higher completely symmetric representations out of the representations of
the $\sltc$ algebras (see Appendix~B) we conclude that the vertex
coordinates of the $k$-th representation of $\slnc$ are scaled by a factor
of $k$ as we go to higher $k$.
Therefore, at least for the vertex points of the $k$-th
representation, we can use the same relation as Eq.~(\ref{eq:beta1}) except
that it is scaled by $k$.

We proceed as with the case $k=1$ and first consider the term
$k|\beta_1|^2\proj{k\ 0\ldots 0}$.  In the root space diagram this is a vertex
of the simplex describing the representation.  We can write the tensor
$\proj{k\ 0\ldots 0}$ as $A^{(11)}$, which is a
$2k$-index tensor with $1$ whenever the $(i_q,j_q)$ index pair is
$(1,1)$ (which happens $k$ times) we see that the boson operator
corresponding to this tensor is
\[ k|\beta_1|^2\sum_{i_1,\ldots,i_k,j_1,\ldots,j_k=1}^d
a_{i_1}^{\dg}\ldots
a_{i_k}^{\dg}A^{(11)}_{i_1,\ldots,i_k,j_1,\ldots,j_k}a_{j_1}\ldots
a_{j_k}.\]
Using the form of $A^{(11)}$ we can simplify this expression and use the
boson algebra to calculate as follows
\begin{equation}\label{beta_k_corner}
k|\beta_1|^2(a_1^{\dg})^k(a_1)^k = \frac{|\beta_1|^2}{d}\Big(k\sum_{i=1}^d
(a_i^{\dg})^k(a_i)^k + k\sum_{j=1}^{d-1}(a_1^{\dg})^k(a_1)^k -
(a_{j+1}^{\dg})^k(a_{j+1}^k)).
\end{equation}
We see that these are exactly the operator analogues of the representation
of the diagonal generators in the higher $k$ representations and we have
the same coefficients as we had in the $k=1$ case.  The terms corresponding
to the other vertices in the root diagram are handled exactly the same way.


We now have to deal with the rest of the diagonal terms which do not
correspond to vertices of the root space diagram.
A typical such (diagonal) term is of the form
\[ (l_{I,i}+1)|\beta_i|^2\proj{I^{(i)}}.\]
Using the transformation from tensors to boson operators given in
Eq.~(\ref{eq:bosrepk}) we get
\[
(k-l)|\beta_i|^2\prod_{i=1}^d(a_i^{\dg})^{k_i}\prod_{i=1}^d(a_i)^{k_i},
\]
where the factor $(k-l)$ is $(l_{I,i}+1)$ where $l$ is the number of steps
along the lattice from the vertex state.  Now we trivially have
\begin{equation}\label{eq:beta_k_inside}
 (k-l)|\beta_i|^2\cA^{[k_i]} = \frac{|\beta_i|^2}{d}\Big(k\cA^{[k_i]} + k(d-1)\cA^{[k_i]} -ld\cA^{[k_i]}),
\end{equation}
where we have used the abbreviation
\[ \cA^{[k_i]} = \prod_{i=1}^d(a_i^{\dg})^{k_i}\prod_{i=1}^d(a_i)^{k_i}.\]

We now argue that these terms are the boson operators corresponding to the
generators of the Lie algebra in the completely symmetric representation.
The states adjacent to the vertex state $\ket{k\,0\dots0}$
are all states of the form $\ket{k-1\dots}$.  Recall that these
states lie on a regular lattice, each segment of which is parallel to
one of the (not necessarily simple) roots.  We first consider the
neighboring state of the form $\ket{k-1\ 1\dots0}$.

Using the standard coordinate system of the root space diagram, we see
that to move to this state means to subtract two from the coordinate
parallel to the segment connecting $\ket{k\,0\dots0}$ and $\ket{k-1\
  1\dots0}$ and one from $(d-2)$ other coordinates, since $1$ is the
projection of the segment on to the remaining $(d-2)$ axes.  To get to the
rest of the states of the form $\ket{k-l\ l\dots0}$, which are further
away, we repeat the procedure $l$ times.
Noting that
$l(-2-(d-2)\times1)=-ld$ we get from Eq.~(\ref{beta_k_corner})
\begin{equation}\label{eq:beta_k}
         {|\b_1|^2\over d}\Big(k\cA^{[k_i]} + k(d-1)\cA^{[k_i]} -ld\cA^{[k_i]}).
\end{equation}
This is exactly what we had in Eq.~(\ref{eq:beta_k_inside}).

The argument for Eq.~(\ref{eq:beta_k}) holds for any state $\ket{k-l\dots}$
lying in the interior of the simplex describing the lattice of points in
the root space diagram.  We first observe that all states of the form
$\ket{k-l\dots}$ lie $l$ segments away from the vertex point
$\ket{k\,0\dots0}$ and thus lie on a single hyperplane and are equidistant
from the state $\ket{k\,0\dots0}$.  To get to any of the states on this
hyperplane from the vertex we always travel $l$ segments of length two.
To reach all these points one does not, of course, traverse segments that
are all pointing in the same direction but each segment is parallel
to one of the coordinates axes of the root space.  Clearly, also in this
case, there will be $(d-2)$ projections of length one to the remaining
(non-parallel) coordinates for each segment.  There may be multiple paths
from a vertex to a given state but this does not affect the argument, which
applies to any of the shortest paths.

We conclude that when going from $k=1$ to $k>1$ the coefficients
$n_\a=|\b_1|^2$ remain the same for the $(d-1)$ diagonal generators from
Eq.~(\ref{eq:allblocks}).  Again the symmetry of the expression shows that
the same argument works with all the $\beta_i$.
\end{proof}
The diagonal terms of $\sigma_A$ are expressed in terms of the $H_{ij}$
diagonal generators of $\slnc$.  The off diagonal terms will correspond to
the step operators of the form $E_{ij}$.  In the previous proof we have
been careful to distinguish between the matrix or tensor describing an
element of a representation and the boson operator corresponding to it.
This correspondence should be clear now and in the next lemma we will just
pretend that the boson operators \emph{are} the generators of the Lie
algebra in the representation.  This will avoid having to insert and remove
hats as we did in the last lemma when we were trying to keep the
distinction clear.
\begin{lem}\label{lem:offdiagk1}
    The off-diagonal part of Eq.~(\ref{eq:allblocks}) is a sum of $\slnc$
    step operators in the $k$-th completely symmetric representation as
    specified in Theorem~\ref{thm:invariant_structure}
    and, for a given $d$, the off-diagonal algebra generator coefficients
    $n_\a$  are  independent of the representation.
\end{lem}
\begin{proof}
    \noindent\textbf{Case} $k=1$.  As we discuss in the appendix
    the step operators of the $\sltc$ subalgebra correspond to
    the edges of complete graphs with states $\ket{0\dots1_l\dots}$
    occupying the vertices.
    Starting at the $l$-th vertex we can see that
    the bosonic step operator from Theorem~\ref{thm:bosrep} takes us to the
    $m$-th vertex
    $$
    a_m^\dg a_l\ket{0\dots1_l\dots}=\ket{0\dots1_m\dots}.
    $$
    Choosing all possible edges (all coefficients $\b_l\bar\b_m$) we can
    fully describe the off-diagonal part of the density matrix
    corresponding to Eq.~(\ref{eq:example_dk1}):
\begin{equation}\label{eq:k1offdiag}
    \sum_{\genfrac{}{}{0pt}{}{m,l=1}{m<l}}^d\b_l\bar\b_m a_m^\dg a_l+h.c.
\end{equation}

    There are $d(d-1)$ summands in total and we see that
    $$
    n_\a=\b_l\bar\b_m,
    $$
    where $\a=(l,m),l\not=m$.
    \begin{rem}
        The raising operators of the form $a_{l+1}^\dg a_l$ precisely
        correspond to the step operators of the fundamental representation
        written in the Chevalley-Serre basis (the $E_{lm}$s) introduced in the appendix.
    \end{rem}
    \noindent\textbf{Case} $k>1$.  Following the strategy from the case
    $k=1$ we choose a particular direction from the $l$-th to the $m$-th
    vertex by setting the coefficients $\b_i\bar\b_j$ where
    $(i,j)\not=(l,m)$ to zero.  Using basic facts from the appendix and
    the observation that to get from one vertex state to the other,
    one has to pass over $k-1$ intermediate states lying on the connecting
    edge. These states divide the edge into $k$ segments. So we have to
    apply the step operator $a_m^\dg a_l$ $k$-times.  Now we consider the
    other lines parallel to the edge under consideration.  The only
    difference is that the number of segments is less than $k$ so the step
    operators have to be applied fewer times.  But these are precisely the
    higher-dimensional step operators written in the bosonic
    representation.

    From Eq.~(\ref{complete_state_traced}) we see that each pair of
    neighboring states (no matter on which line parallel to the
    edge under consideration they lie) have the following coefficient
    $$
    \b_l\bar\b_m\sqrt{(l_{I,l}+1)(l_{I,m}+1)}\kb{I^{(l)}}{I^{(m)}}.
    $$
    However, using the rules for bosonic creation and annihilation
    operators we find that
    \begin{equation}\label{eq:koffdiag}
        a_m^\dg a_l\ket{I^{(l)}}=\sqrt{l_{I,l}+1}\sqrt{l_{I,m}+1}\ket{I^{(m)}}.
    \end{equation}
    We again conclude that in the transition from $k=1$ to $k>1$ the
    expansion coefficients in Eq.~(\ref{eq:allblocks}) remain identical
    $$
    n_\a=\b_l\bar\b_m,
    $$
    where $\a=(l,m),l\not=m$.
\end{proof}
\begin{proof}[Proof of Theorem~\ref{thm:invariant_structure}]
    Putting together Lemmas~\ref{lem:diagk1} and~\ref{lem:offdiagk1}
    capturing separately the diagonal and off-diagonal parts of
    Eq.~(\ref{eq:allblocks}) gives the theorem statement. The total number of operators is indeed $L=2d(d-1)$ where
    $d(d-1)$ of them comes from the diagonal part (see Eqs.~(\ref{eq:beta1op}) and (\ref{beta_k_corner}) and note that we repeat the procedure $d$-times). The remaining $d(d-1)$ operators come from the off-diagonal part (Eq.~(\ref{eq:k1offdiag}) for $k=1$). For $k>1$ the only difference is that we act with all $d(d-1)$ operators from Eq.~(\ref{eq:k1offdiag}) $k$-times as witnessed in Eq.~(\ref{eq:koffdiag}).
\end{proof}
\begin{rem}
  In the last paragraph of Appendix \ref{app:geom.repn}, we discuss why the
  generators constructed above are indeed generators of the $k$-th
  completely symmetric representation of $\slnc$, in particular, why they
  satisfy the necessary commutation relations.
\end{rem}

We have seen that the Hilbert spaces carry representations of
$SU(d)$; we now show that the transformations effected by the Unruh channel
mesh properly with the group actions on the state spaces: the qudit Unruh
channel is $SU(d)$-\emph{covariant}.  This property is very helpful in evaluating is
regular and private quantum capacities.
Recall that $\dm{(\H)}$ stands for the
space of density matrices on the Hilbert space $\H$.
\begin{defi}
Let $G$ be a group, $\H_{in},\H_{out}$ be Hilbert spaces and let $r_1:G\to
GL(\H_{in}),r_2:G\to GL(\H_{out})$ be two unitary representations of the
group $G$.  Let $\K:\dm\big(\H_{in}\big)\to\dm\big(\H_{out}\big)$ be a channel.
We say that $\K$ is \textbf{covariant with respect to} $G$ and the
representations $r_1,r_2$, if
\begin{equation}\label{eq:covcond}
    \K\left(r_1(g)\rho r_1(g)^\dg\right)=r_2(g)\K(\rho)r_2(g)^\dg
\end{equation}
holds for all $g\in G,\rho\in\dm(\H_{in})$.
\end{defi}

The covariance of the qudit Unruh channel is an easy consequence of the main structure theorem.
\begin{cor}\label{cor:Covariance-Unruh}
    The qudit Unruh channel is $SU(d)$-covariant for the fundamental
    representation in the input space and any of the completely symmetric
    representations carried by the output space of the channel.
\end{cor}
\begin{proof}

The channel output $\s_A$ in Eq.~(\ref{eq:Unruhchanneloutput}) is an
infinite-dimensional block-diagonal trace class matrix. It can be rewritten
as
\begin{equation}\label{eq:Unruhchanneloutput_rewritten}
    \s_A=\bigoplus_{k=1}^\infty s_k\tilde\s_A^{(k)},
\end{equation}
where $s_k=(1-z)^{d+1}z^{k-1}{d+k-1\choose d}$. $\tilde\s_A^{(k)}$ is proportional to $\s_A^{(k)}$ such that
$\Tr{}\tilde\s_A^{(k)}=1$ for all $k$.  This implies that
the qudit Unruh channel can be written as
$\E(\psi_A)=\bigoplus_{k=1}^\infty s_k\E_k(\psi_A)$, where the
$\E_k(\psi_A)$ can be read off Eq.~(\ref{eq:Unruhchanneloutput}).
Suppose that $\psi_A \mapsto \psi'_A = r_1(g) \psi_A r_1(g)^\dg$ for $r_1$ the defining representation of $SU(d)$ on the multi-rail qudit encoding space. Then
\begin{equation} \label{eqn:covar1}
	\psi_A
	={1\over d}\Big(\openone+\sum_{\a=1}^{L}n_\a\lam^{(1)}_\a\Big)
	\quad \mbox{and} \quad
	\psi_A'
	={1\over d}\Big(\openone+\sum_{\a=1}^{L}n'_\a\lam^{(1)}_\a\Big)
\end{equation}
for some choices of $n_\a$ and $n_\a'$. Consider the map taking
\begin{equation} \label{eqn:covar2}
    \s_A^{(k)}={1\over d}\Big(k\openone+\sum_{\a=1}^{L}n_\a\lam^{(k)}_\a\Big)
    \quad \mbox{to} \quad
    \s_A'^{(k)}={1\over d}\Big(k\openone+\sum_{\a=1}^{L}n_\a'\lam^{(k)}_\a\Big).
\end{equation}
First we will verify that this map is well-defined despite the fact that the choices of $n_\alpha$ and $n_\alpha'$ are not unique. If $\sum_\a n_\a \lambda_\a^{(1)} = \sum_\a m_\a \lambda_\a^{(1)}$, then this identifies a linear relation in the
Lie algebra $\slnc$ because the $\lambda_\a^{(1)}$ are elements of the
fundamental representation of $\slnc$. Any such linear relation must also hold in any representations of $\slnc$, in particular, the $k$th completely symmetric representations. It follows that $\sum_\a n_\a \lambda_\a^{(k)} = \sum_\a m_\a \lambda_\a^{(k)}$ so the map is well-defined and, in fact, linear.

The only linear map satisfying Eq.~(\ref{eqn:covar2}), however, is conjugation by $r_2(g)$ where $r_2$ is $k$-fold symmetric power of $r_1$, so we have verified that the following diagram commutes:
\[
\xymatrix{
\psi_A\ar[r]^{\E_k}\ar[d]_{r_1(g)} & \tilde\s_A^{(k)}\ar[d]^{r_2(g)}\\
\psi_A'\ar[r]_{\E_k}& \tilde\s_A'^{(k)}\\
}
\]

Therefore, the covariance condition Eq.~(\ref{eq:covcond}) holds for all
$\E_k$.  Since the output of the qudit Unruh channel is a direct sum of
$\E_k(\psi_A)$ we conclude that the qudit Unruh channel is covariant as well.
\end{proof}
\begin{rem}
  We emphasize that the covariance proof shows only that the restriction of
  the Unruh channel to the input space spanned by input qudit states from
  Eq.~(\ref{eq:multi-railbasis}) is covariant. As a matter of fact, one can
  easily show that for a general input state the Unruh channel is not
  covariant with respect to $SU(d)$.
\end{rem}

\section{Quantum capacities of the qudit Unruh channel}
\label{sec:capcalc}

While there is no known single-letter formula for the quantum
capacity of a general quantum channel, if a channel has the property
of being either degradable or conjugate degradable, the optimized
coherent information does give such a formula~\cite{devetak2005degr,conjdeg}. It was
shown in~\cite{qubitunruh} that the qubit Unruh channel is conjugate
degradable. We will show below that this property extends to the
qudit Unruh channels. From there, we will calculate the quantum
capacity.
\begin{defi}
A channel $\E$ is \textbf{conjugate degradable} if there exists a
quantum channel $\Dcd$, called a \textbf{conjugate degrading map},
which degrades the channel to its complementary channel $\E_c$ up to
complex conjugation $\C$:
\begin{equation}\label{def:conjdeg}
    \Dcd\circ\E=\C\circ\E_c.
\end{equation}
\end{defi}
\begin{thm}\label{thm:cd_of_UnruhSUdchannel}
    The qudit Unruh channel $\E$ from Alice to Eve introduced in
    Def.~\ref{def:Unruhchannel} is conjugate degradable. The explicit
    transformation to the complementary output is
    \begin{equation}\label{conjdegmap}
    \E_c(\psi_A) = z\bar\sigma_A + (1 - z )\omega_0,
    \end{equation}
    where $\sigma_A=\E(\psi_A)$ and $\omega_0$ is a diagonal state
    independent of $\sigma_A$.
\end{thm}
The proof of the theorem will be preceded by two lemmas for which
purpose we rewrite Eq.~(\ref{eq:transformed_qudit_final}) as
    \begin{equation}\label{eq:transformed_qudit_final_rewritten}
    \ket{\s}_{AC}=(1-z)^{(d+1)/2}\sum_{k=1}^\infty z^{(k-1)/2}\ket{\s^{(k)}}_{AC}.
    \end{equation}
\begin{lem}\label{lem:conjdeg}
    The following relation holds: $\s_C^{(2)}=\bar\s_A^{(1)}+\openone$
    where $\s_C^{(k)}=\Tr{A}\s^{(k)}_{AC}$ and $\s_A^{(k)}=\Tr{C}\s^{(k)}_{AC}$.
\end{lem}
\begin{proof}
    We rewrite a part of the state
    Eq.~(\ref{eq:transformed_qudit_final_rewritten}), namely
    $\ket{\s^{(2)}}_{AC}$, as
    \begin{equation}\label{transformed_qudit_final_LEXI_rev}
        \ket{\s^{(2)}}_{AC}=\sqrt{2}\sum_{i=1}^d\beta_i\ket{ii}_A\ket{i}_C
        +\sum^{\binom{d}{2}}_{\ahead{i,j}{i\not=j}}\ket{ij}_A\left(\beta_j\ket{i}+\beta_i\ket{j}\right)_C,
    \end{equation}
    where for the $A$ subsystem $\ket{ii}_A$ labels a $d$-mode Fock state
    where the $i$-th position is occupied by two photons.  $\ket{ij}_A$
    labels a $d$-mode Fock state where the $i$-th and $j$-th positions are
    occupied by single photons. There are no other possibilities. The
    $C$~subsystem is even simpler since $\ket{i}_C$ just means the $i$-th
    position being occupied by a single photon. This labeling has the
    advantage of having the same form for all $d$. Tracing over the
    $A$ subsystem we get
    \begin{equation}
        \s_C^{(2)}=\sum_{i=1}^d\left(2|\beta_i|^2+\sum^{d-1}_{j\not=i}|\beta_j|^2\right)\kb{i}{i}
        +\sum^{\binom{d}{2}}_{\ahead{i,j}{i\not=j}}\beta_j\bar\beta_i\kb{i}{j}.
    \end{equation}
    Applying the normalization condition $\sum_{i=1}^d|\beta_i|^2=1$ we find
    \begin{equation}\label{eq:complementoutput_fork2}
        \s_C^{(2)}=\openone+\sum_{i=1}^d|\beta_i|^2\kb{i}{i}+\sum^{\binom{d}{2}}_{\ahead{i,j}{i\not=j}}\beta_j\bar\beta_i\kb{i}{j}.
    \end{equation}
    Expanding Eq.~(\ref{eq:transformed_qudit_final_rewritten}) for $k=1$ then gives
    \begin{equation}
         \ket{\s^{(1)}}_{AC}=\sum_{i=1}^d\beta_i\ket{i}_A\ket{0\dots0}_C.
    \end{equation}
    (We are abusing a notation a bit by mixing both ket conventions.) By tracing over the $C$~subsystem we get
    \begin{equation}\label{output_fork1}
        \s_A^{(1)}=\sum_{i=1}^d|\beta_i|^2\kb{i}{i}+\sum^{\binom{d}{2}}_{\ahead{i,j}{i\not=j}}\beta_i\bar\beta_j\kb{i}{j}.
    \end{equation}
    Comparing with Eq.~(\ref{eq:complementoutput_fork2}) completes the proof.
\end{proof}
    This shows that, at least for $k=2$, the complementary output is complex conjugated and
    admixed with a maximally mixed state with respect to some part of the
    qudit Unruh channel output. Equally importantly, we see that
    $\s_C^{(2)}$ has an algebra generator structure closely related
    to that of $\s_A^{(1)}$.
\begin{lem}\label{lem:repre_Change}
    The following relation holds for all $k$:
    $\s_C^{(k+1)}=\bar\s_A^{(k)}+\openone$.
\end{lem}
\begin{proof}
In Theorem~\ref{thm:invariant_structure} we explicitly showed that
even as the size of the underlying representation is increased, the expansion
coefficients of $\sigma_A$ in terms of the Lie algebra generators can be chosen to
be independent of the representation. The same is actually true of the $C$ subsystem.
If we simply rewrite  the core of
Eq.~(\ref{Taylor_summands})
\begin{equation}\label{Taylor_summands_rewritten}
    \left\{{1\over\sqrt{l_1!\dots l_d!}}\bigotimes_{i=1}^d\big(a_i^\dg\big)^{l_i}
    \otimes{1\over\sqrt{l_1!\dots l_d!}}\bigotimes_{i=1}^d\big(c_i^\dg\big)^{l_i}\right\}_{\sum
    l_i+1=k}
\end{equation}
then the left product composed of $a_i^\dg$ operators generates the
$A$ subsystem whose structure has been completely described. But the
right product is identical to the left one and so
Theorem~\ref{thm:invariant_structure} is applicable for the
$C$~subsystem as well. In other words, taking
Eq.~(\ref{eq:complementoutput_fork2})  we know exactly how any other
$\s_C^{(k+1)}$ will look like and we may conclude that
$$
\s_C^{(k+1)}-\openone=\bar\s_A^{(k)}.
$$
\end{proof}
\begin{proof}[Proof of Theorem~\ref{thm:cd_of_UnruhSUdchannel}]
    Let us explicitly construct the conjugate degrading map, which isn't
    hard given
    the relationships we've identified between the output of the qudit Unruh
    channel and the output of its complementary channel.
    \begin{align}
      \s_A &= (1-z)^{d+1}\left[\s_A^{(1)}\oplus z\s_A^{(2)}\oplus
        z^2\s_A^{(3)}\oplus\dots\right] \\
      \s_C &= (1-z)^{d+1}\left[\kb{0\dots0}{0\dots0}_C\oplus
        z\s_C^{(2)}\oplus z^2\s_C^{(3)}\oplus\dots\right].
    \end{align}
    We admix the complex conjugated $\s_A$ with a properly chosen diagonal
    state and use Lemma~\ref{lem:repre_Change} to get
    \begin{equation}\label{eq:conjdegmap}
        \s_C = z\bar\s_A+(1-z)\o_0,
    \end{equation}
    where $\o_0=(1-z)^d\left[\kb{0\dots0}{0\dots0}\oplus z\openone\oplus
      z^2\openone\oplus\dots\right]$. This concludes the proof.
\end{proof}

If a channel is covariant and conjugate degradable, then the maximization in the formula for the
quantum capacity from Theorem~\ref{thm:lsd} is achieved with a
maximally mixed input qudit $\pi_A$. (See the calculation leading up to Eq.~(9) in \cite{conjdeg}.) The same happens for the evaluation of the formula for the private quantum capacity in Theorem~\ref{thm:qpriv}. Since we have shown that the qudit Unruh channel is both covariant and conjugate degradable, we must therefore
calculate $\E(\pi_A)$ and $\E_c(\pi_A)$.

The image of a single input pure state, say $\ket{1}$, reads
\begin{equation}\label{eq:diagstatetransformation}
    \E:\ket{1}_A\mapsto(1-z)^{d+1}\bigoplus_{k=1}^\infty
    z^{k-1}\ \sum_{I^{(1)}}(l_{I,1}+1)\kb{I^{(1)}}{I^{(1)}}_A,
\end{equation}
where we recall that $(l_{I,1}+1)$ is the first label of
$\ket{I^{(1)}}_A$ for a given $k$ and $d$. Let $\pi_{1 \to i}$ be the permutation transposing $1$ and $i$. For the input ket $\ket{i}=\pi_{1\to i}\ket{1}$
we get from Eq.~(\ref{eq:beta1})
\begin{equation}\label{eq:diagstatetransformation_i}
    \E:\ket{i}_A\mapsto(1-z)^{d+1}\bigoplus_{k=1}^\infty
    z^{k-1}\ \sum_{I^{(i)}}(l_{I,i}+1)\kb{I^{(i)}}{I^{(i)}}_A.
\end{equation}
Since $\sum_{i=1}^d(l_{I,i}+1)=k$ and $\pi_A = 1/d \sum_{i=1}^d \kb{i}{i}_A$ we get
\begin{equation}\label{eq:maxmixedsigma}
    \E:\pi_A\mapsto\rho_A={1\over d}(1-z)^{d+1}\bigoplus_{k=1}^\infty kz^{k-1}
    \sum_{I^{(i)}=1}^{d+k-1\choose k}\kb{I^{(i)}}{I^{(i)}}_A,
\end{equation}
where the sum over $i$ from $\pi_A$ is hidden in the sum over $I^{(i)}$. The example from Eq.~(\ref{eq:exampled3k3}) can be helpful. Observe that, on each irrep, the state is proportional to the identity as
required by Schur's lemma.
\begin{figure}[t]
\begin{center}
    \resizebox{14cm}{6cm}{\includegraphics{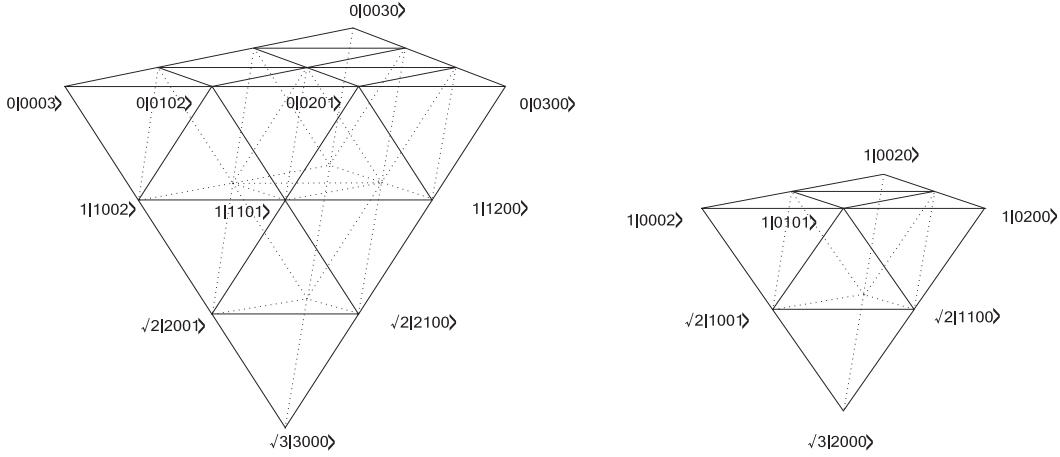}}
    \caption{We illustrate the calculation of the image of a maximally input mixed input qudit $\rho_A = \E(\pi_A)$ (on the left) and $\rho_C = \E_c(\pi_A)$ (on the right) on $d=4$ and $k=3$. Both plots capture the situation where $\b_1=1$. We can visualize the calculations in Eqs.~(\ref{eq:maxmixedsigma}) and~(\ref{eq:maxmixedsigma_Compl}) if we realize that the permutation $\pi_{1\to i}$ preserves the coefficients $\sqrt{k}$ when acting on $\ket{k\,0\dots}$ (on the left) or on $\ket{k-1\dots}$ (on the right). The summing procedure $\sum_{i=1}^d \kb{i}{i}_A$ is like adding $d$ rotated simplices with squared coefficients.}
    \label{fig:tetrahedron}
\end{center}
\end{figure}
\begin{figure}[h]
\begin{center}
    \resizebox{11cm}{8cm}{\includegraphics{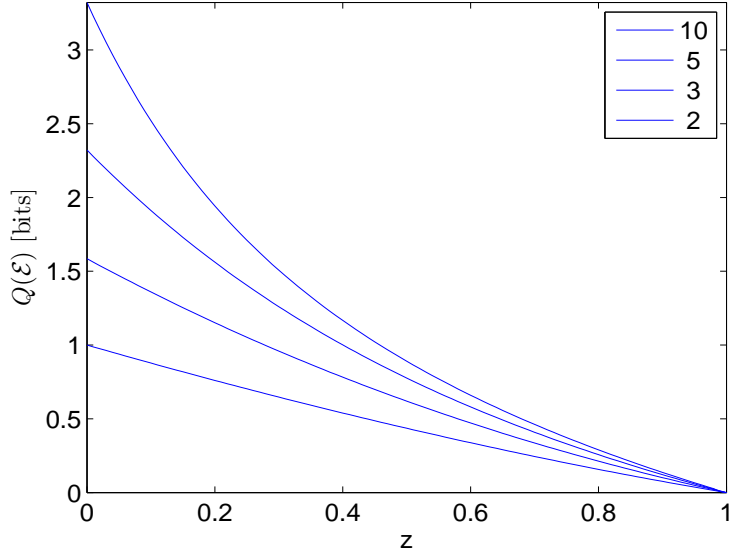}}
    \caption{The  quantum capacity as calculated by
      Eq.~(\ref{eq:quantcap_calculation}) for several qudit Unruh channels.
      The curve achieving a capacity of 1 for $z=0$ corresponds to $d=2$. The others, in order of increasing quantum capacity, are $d=3, 5$ and $10$.}
    \label{fig:Q_bits}
\end{center}
\end{figure}
\begin{figure}[h]
\begin{center}
    \resizebox{11cm}{8cm}{\includegraphics{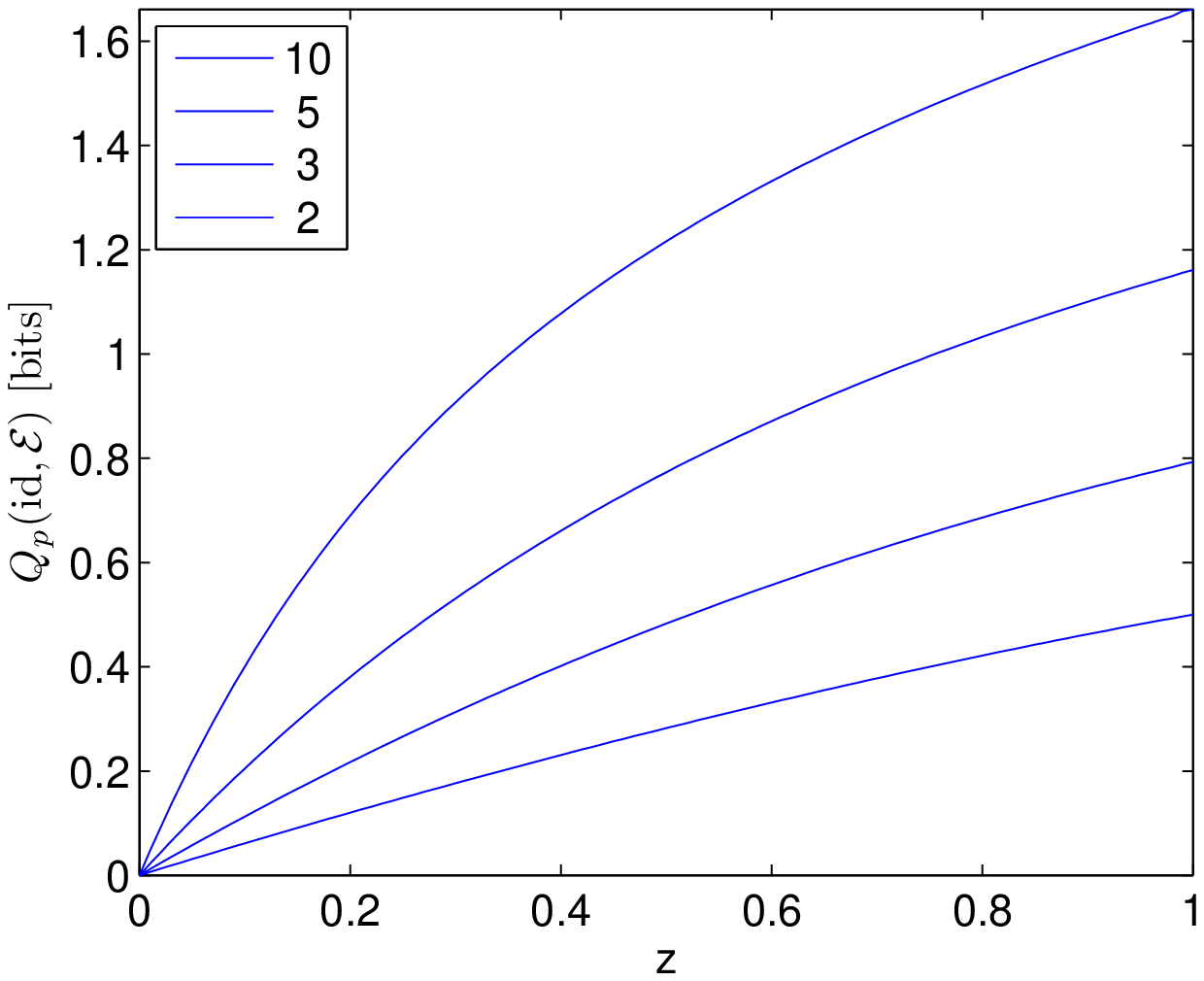}}
    \caption{The private quantum capacity as calculated by Eq.~(\ref{eq:privcap_Calculation}) for several qudit Unruh channels. In order of increasing capacity, the curves correspond to $d=2,3,5$ and $10$.}
    \label{fig:Qp_bits}
\end{center}
\end{figure}
The output from the channel complementary to the Unruh channel will also be needed. Once we know the structure of the $k$-th output block of the Unruh channel  Lemma~\ref{lem:repre_Change}, tells us about the structure of the $(k+1)$-th block of its complementary channel. Its dimension must be same and so the states $\ket{I}_C$ span the ${d+k-2\choose k-1}$-dimensional completely symmetric subspace of $(k+1)$~photons. But we prefer to compare the $k$-th complementary block with the $k$-th Unruh channel output block because in our notation $\ket{I}_C$ contains one photon less than $\ket{I^{(i)}}_A$. It follows from Lemma~\ref{lem:repre_Change} and Eq.~(\ref{eq:diagstatetransformation}) that
\begin{equation}\label{diagstatetransformation_Complement}
    \E_c:\ket{1}_A\mapsto(1-z)^{d+1}\bigoplus_{k=1}^\infty
    z^{k-1}\ \sum_{I=1}^{d+k-2\choose k-1}(l_{I,1}+1)\kb{I}{I}_C.
\end{equation}
for $\b_1=1$ and as a result the coefficients $(l_{I,1}+1)$ remain the same. Because the eigenvalues are equal we see that the $k$-th block of the complementary channel is the complement of the $k$-th block of the Unruh channel. Fig.~\ref{fig:tetrahedron} shows an example of the relation between two complementary blocks. Since for $\b_1$ the coefficient $(l_{I,1}+1)$ for the `lowest' states $\ket{0\,k\,0\dots}$ equals one, the action of $\pi_{1\to i}$ transfers the coefficient one to the `highest' state $\ket{k\,0\dots}$. This happens exactly $(d-1)$ times (that is, for all remaining $\b_i$'s, $i=2\dots d$) and so we may conclude that a maximally mixed input qudit transforms as
\begin{equation}\label{eq:maxmixedsigma_Compl}
   \E_c:\pi_A \mapsto\rho_C={1\over d}(1-z)^{d+1}\bigoplus_{k=1}^\infty
    (d+k-1)z^{k-1}\ \sum_{I=1}^{{d+k-2\choose k-1}}\kb{I}{I}_C.
\end{equation}
Once again, the state is proportional to the identity on each irrep. We
will take Eq.~(\ref{eq:maxmixedsigma_Compl}) as the definition of
$\rho_C$ for the remainder of the paper.
\begin{figure}[h]
\begin{center}
    \resizebox{11cm}{8cm}{\includegraphics{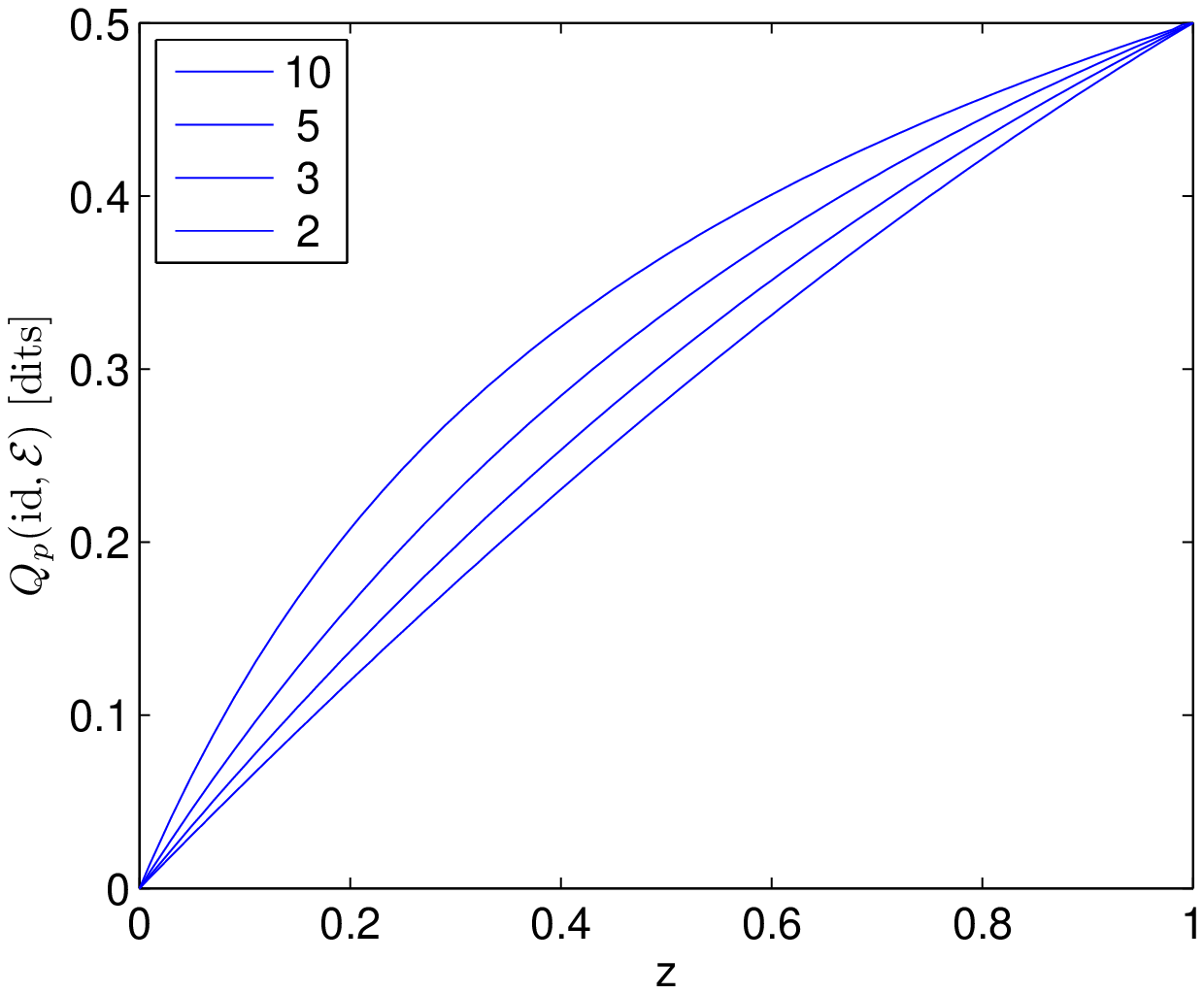}}
    \caption{The private quantum capacity given by Eq.~(\ref{eq:privcap_Calculation}) for several qudit Unruh channels in units of dits. The uppermost curve corresponds to $d=10$, then in order $d=5,3,2$. This presentation facilitates comparison of the private quantum capacity for different $d$ by correcting for the fact that a noiseless qudit channel can send $\log d$ times as much information as a noiseless qubit channel. In these units, one immediately sees that in the limit of infinite acceleration, the private quantum capacity approaches a value of $\smfrac{1}{2} \log d$, meaning that Alice and Bob need only sacrifice half of their transmission bandwidth to secure their messages. More interestingly, the graph indicates that using higher $d$ yields more efficient encodings for finite values of Eve's acceleration.}
    \label{fig:Qp_dits}
\end{center}
\end{figure}
We define $T_{d,z}=1/d(1-z)^{d+1}$ and after some straightforward
algebra we get
\begin{equation}\label{entropy_kappad}
    H(A)_\rho=-\log T_{d,z}-(1+d){z\over1-z}\log z-T_{d,z}\sum_{k=1}^\infty{d+k-1\choose k}kz^{k-1}\log k.
\end{equation}
Similarly, for the complementary
output~Eq.~(\ref{eq:maxmixedsigma_Compl})
\begin{align}\label{entropy_kappad_Compl}
    H(C)_\rho&=-\log T_{d,z}-(1+d){z\over1-z}\log z\nn\\
             &-T_{d,z}\sum_{k=1}^\infty{d+k-2\choose k-1}(d+k-1)z^{k-1}\log{(d+k-1)}.
\end{align}
The quantum capacity of the qudit Unruh channel simplifies to
\begin{equation}\label{eq:quantcap_calculation}
Q(\E)=H(A)_\rho-H(C)_\rho=T_{d,z}\sum_{k=1}^\infty{d+k-1\choose k}kz^{k-1}\log{d+k-1\over k}.
\end{equation}
We present the plot of the quantum capacity as a function of the acceleration parameter in Fig.~\ref{fig:Q_bits}.

For the private quantum capacity we recall our single-letter formula from
Theorem~\ref{thm:qpriv}. The channel to Bob is a noiseless channel
and so
\begin{align}\label{eq:privcap_Calculation}
Q_p(\id,\E)&=\smfrac{1}{2} I(A';C)_\rho\nn\\
&=\smfrac{1}{2}\big(\log{d}+H(C)_\rho- H(A)_\rho\big)\nn\\
&=\smfrac{1}{2}\Bigg(\log{d}-T_{d,z}\sum_{k=1}^\infty
{d+k-1\choose k}kz^{k-1}\log{d+k-1\over k}\Bigg).
\end{align}
The private quantum capacity is plotted in Figs.~\ref{fig:Qp_bits} and~\ref{fig:Qp_dits}. The second figure demonstrates that private communication is more efficient with qudit encodings than with qubit encodings even after normalization for the fact that a qudit channel carries more information than a qubit channel when $d>2$.
Therefore, for the qudit Unruh channel more efficient encodings are possible.

\section{Conclusions}

We investigated two communication problems in Rindler spacetime. The
first was to determine the optimal rate at which a sender could
reliably transmit qubits to a uniformly accelerating receiver. While
this problem has resisted solution for general quantum channels, in
the case of the qudit Unruh channels, we are able to extract a
compact, tractable formula which is strictly positive for all
accelerations. In order to evaluate the capacity, we decomposed the
output of the Unruh channel into irreducible completely symmetric
representations of the unitary group. From this decomposition,
we were able to show that the channels have a rare and useful
property known as conjugate degradability, which makes the
calculation of the capacity possible.

The second problem involves securely sending encrypted quantum
information from an inertial sender to an inertial receiver in the
presence of an accelerating eavesdropper. Because the associated
general private quantum capacity problem had only been very briefly
discussed previously, we began by studying it for arbitrary
channels. In the case where the channel from the sender to the
intended receiver is noiseless, our formula ``single-letterizes'',
meaning that it involves no intractable limits. Specifically, the
private quantum capacity is equal to the entanglement-assisted
capacity to the eavesdropper's environment. Applied to the qudit
Unruh channels, we find the private quantum capacity is positive
for all non-zero eavesdropper accelerations, no matter how small.

While we have phrased all our results in the language of Rindler
spacetime and accelerating observers, the mathematics also describes
the noise induced by a nonlinear optical parametric amplifier (NOPA)~\cite{braunstein2001optimal,caves}. Our quantum capacity result therefore indicates that the quantum capacity through such an
amplifier with arbitrarily high gain is always strictly positive and
can be exactly calculated.

A natural direction for future study would be to relax some of the
assumptions made in this article. First, it would be more natural to
impose a power restriction, in the form of the average number of
photons per channel use, than to restrict to the $d$-rail encodings
we study here~\cite{holevo2008}. It would also be interesting to use a more realistic
model of the channel from sender to receiver than the noiseless
channel studied here. Finally, we have been very conservative in
modeling the eavesdropper, allowing her to perform arbitrary
operations on her Rindler modes, ignoring her necessarily finite
extent. While moving to a power restriction is unlikely to change
the qualitative features of our conclusions, there is significant
room for new effects when studying realistic receiver and
eavesdropper channels. In particular, the quantum capacity would
likely vanish at a finite acceleration and the private quantum
capacity might only be non-zero for sufficiently high accelerations.

\section*{Acknowledgements}

We would like to thank Keshav Dasgupta, Alex Maloney and Mark Wilde
for helpful discussions.  This work was supported by a grant from the
Office of Naval Research (N000140811249).  We also gratefully
acknowledge the support of the Canada Research Chairs program,
CIFAR, INTRIQ, MITACS, NSERC, the Perimeter Institute and
QuantumWorks.

\appendix
\section{Background on representation theory}
\label{appendix}

In the main body of the paper we exploited the covariance of the Unruh
channel in order to calculate quantities of interest.  This used some
standard material on the representation theory of Lie algebras that may not
be familiar to all readers.  In this section we collect the relevant
definitions for the benefit of such a reader.

The representations of any Lie group are closely related to the
representations of the corresponding Lie algebra: in physicists' language
this amounts to working with the ``infinitesimal generators'' of the group.
Mathematically, a Lie group is a group that is also a smooth manifold with
all the group operations being smooth (infinitely differentiable).  The Lie
algebra is the tangent space at the identity.  An easy, but fundamental,
result says that associated with any representation of a Lie group is a
unique corresponding representation of the Lie algebra and the
representation of the Lie group is irreducible if and only if the
corresponding Lie algebra representation is irreducible.  From the
representations of the Lie algebra we can reconstruct the representations
of the connected component of the identity of the Lie group; so, in
particular, if the Lie group is connected the Lie algebra representations
determine the Lie group representations.

The Lie algebra of $SU(d)$ is $\mathsf{su}(d)$ and consists of the complex
self-adjoint\footnote{We use the convention that the passage from the Lie
  algebra to the Lie group is $X\mapsto e^iX$ rather than $X\mapsto e^{X}$,
  the latter is common in the pure mathematics literature.}  matrices with
trace zero.  It is more convenient to work with the complexified form which
is $\mathsf{su}(d)\otimes\mathbb{C}$.  It is easy to see that this is
isomorphic to $\slnc$.  Thus we have to classify the representations of
$\slnc$.

The paradigmatic example of the classification of the finite-dimensional
representations of a Lie algebra is the case of $\sltc$.  Here there are
three generators of the Lie algebra: $J_x,J_y,J_z$, none of them commute
with each other but they all commute with $\mathbf{J}^2=J_x^2+J_y^2+J_z^2$.
The irreps are eigenspaces of $\mathbf{J}^2$ and are usually labelled by
the corresponding eigenvalue of $\mathbf{J}^2$, or more precisely, by a
number that determines the eigenvalue.  In this case the label of each
irrep is a positive integer $j$, which is the dimension of the irrep and
the eigenvalue is $\frac14 (j^2-1)$.  A basis for the irrep is given by the
eigenvectors of $J_z$ and the combinations $J^{\pm} = J_x\pm iJ_y$ act as
raising and lowering operators.  All this is assumed familiar to the
reader; the $\slnc$ case generalizes this situation.

In the $\slnc$ case there may be several mutually commuting operators
instead of just one as in $\sltc$.  A
\emph{maximal} commuting set of operators of a (semi-simple) Lie
algebra\footnote{This is not the right definition for general Lie algebras
  but it is adequate for semi-simple Lie algebras.} is called a
\emph{Cartan subalgebra}.   The Cartan subalgebra of
$\slnc$ has dimension $r=d-1$; we say that the rank of the Lie algebra is
$r$.  We write $\mathbf{H}=(H_1,\ldots,H_r)$ for the Cartan subalgebra
generated by the elements $\{H_1,\ldots,H_r\}$ of the Lie algebra; these
elements are assumed to be linearly independent.
Once a Cartan subalgebra has been chosen we can
use the common eigenvectors of the members of the Cartan subalgebra as
the basis vectors of an irreducible representation just as we used the
eigenvectors of $J_z$ in the case of $\sltc$.

\begin{defi}
An $r$-tuple $\mathbf{\alpha} =(\a_1,\ldots,\a_r)$ of complex numbers is
called a \textbf{root} if: (i) not all the $\a_i$ are zero, (ii) there is an
element $E$ of $\slnc$ such that
\begin{equation}\label{eq:cartanweyl}
[H_i,E]=\a_iE.
\end{equation}
\end{defi}
Typically, we use the root to label $E$, thus, we would write
$[H_i,E_\a]=\a_iE_\a$.  A maximally independent set of the form
$\{H_i,E_\a\}$, where the $H_i$ are a basis of the Cartan subalgebra and
the $E_\a$ are associated with roots as above, is called a Cartan-Weyl
basis~\cite{gilmore,fuchs} of the Lie algebra.

Among all the roots one can choose (nonuniquely) a class of special roots
called positive roots.
\begin{defi}
A set of roots is called a collection of \textbf{positive roots} if: (i) for
any root $\alpha$ either $\alpha$ or $-\alpha$, but not both are in the
set, and (ii) for any
roots $\alpha,\beta$ in the set then if $\alpha + \beta$ is a root it must
also be in the set.
\end{defi}
Once we have designated a family of roots as positive roots we can
define what it means for a root to be simple.
\begin{defi}
A positive root is called a \textbf{positive simple root} if it cannot be
written as a linear combination of other positive roots.
\end{defi}
\begin{defi}
  If $\rho$ (where $\rho:\slnc\to GL(V)$ for some $V$) is a representation
  of $\slnc$ then a $r$-tuple $\mathbf{\mu} =(\mu_1,\ldots,\mu_r)$ of
  complex numbers is a \textbf{weight} for $\rho$ if there is a nonzero
  vector $\psi\in V$, called a \textbf{weight vector}, such that $\psi$ is
  an eigenvector of each $H_i$ with eigenvalue $\mu_i$.
\end{defi}
If $\mathbf{\mu}$ is a weight and $\psi$ a weight vector for $\rho$ and
$\mathbf{\alpha}$ is a root then
\begin{equation}\label{eq:weighteq}
\rho(H_i)\rho(E)\psi = (\mu_i + \a_i) \rho(E)\psi.
\end{equation}

In short, $E$ produces a new weight vector and changes some, perhaps all,
of the eigenvalues of the Cartan operators.  The root is a vector in the
weight space that points in the direction in which the weights are
changing.  The $E$ operators are called shift or raising and lowering
operators; they are the analogues of the $J^{\pm}$ except now there are
many of them pointing in different directions.  Roughly speaking, the
positive roots correspond to raising operators and the negative roots to
lowering operators.  We classify the irreducible representations by the
\emph{highest possible value of the weight}, although doing so of course
requires defining a suitable order on the weights.

A special example of a Cartan-Weyl basis is the Chevalley-Serre
basis~\cite{gilmore,fuchs}.  Two aspects make this basis special: (i) The
step operators are associated to simple roots and (ii) the normalization is
chosen such that the roots are integers. Unless explicitly stated we work
in this basis because it has a useful geometric interpretation.

To give the operators a geometric interpretation we will work with a
specific matrix representation of the $\slnc$ algebra.  We define
$E_{ij}$, where $1\leq i\not=j\leq d$, as the matrix having one where the
$i$-th row and the $j$-th column intersect and the rest of the entries are
zeros.  Furthermore we define a diagonal matrix $H_{ij}$ in which the $i$-th
diagonal entry is 1, the $j$-th diagonal entry is $-1$ and the rest are
zeros.  If we assume $j=i+1$ the set $\{H_{ij},E_{ij},E^\dg_{ij}\}$
forms the Chevalley-Serre basis for $\slnc$.  More explicitly, for $d=2$ we
get
\begin{subequations}\label{eq:sl2C}
\begin{align}
    H_{12}&=\begin{pmatrix}
           1 & 0 \\
           0 & -1 \\
         \end{pmatrix}\\
    E_{12}&=\begin{pmatrix}
             0 & 1 \\
             0 & 0 \\
           \end{pmatrix}\\
    E^\dg_{12}&=\begin{pmatrix}
          0 & 0 \\
          1 & 0 \\
    \end{pmatrix}.
\end{align}
\end{subequations}
It is often the case that one is working with the diagonal matrix
$h_{i(i+1)}$ so we often just write $H_i$ for this.
For the matrices just defined we have the following commutators:
\begin{subequations}\label{eq:sl2cinChevalley}
\begin{align}
    [H_i,E_{ij}]&=2E_{ij}\\
    [H_i,E^\dg_{ij}]&=-2E^\dg_{ij}.
\end{align}
\end{subequations}
This choice of basis opens the way to a geometric picture based on
the so-called root space diagram.

The simple roots are elements of a vector space dual to the one
spanned by elements of the Cartan subalgebra. A root $\alpha$
defines a linear map, also written $\alpha$, on the Cartan
subalgebra by the simple rule $\alpha(H_i) = \alpha_i$; since the
$H_i$ form a basis for the Cartan subalgebra this gives a linear
map.  This definition has the property that for a generic $H$ in the
Cartan subalgebra and $E$ associated with the root $\alpha$ we have
$[H,E] = \alpha(H)E$. Thus we can think of roots as tuples or as
elements of the dual space.

This justifies calling the dual space the space of roots.  For the
$\slnc$ Lie algebra the space of roots is $(d-1)$-dimensional and
the simple root vectors, defined as $r$-tuples of simple roots, form
a basis (which, in general, is not orthogonal).
It can be shown that two consecutive simple root vectors subtend the angle
$2\pi/3$.

In terms of the Chevalley-Serre basis we can write explicit versions
of Eq.~(\ref{eq:cartanweyl}) and Eq.~(\ref{eq:weighteq}).  If we
write $\mu = (\mu_1,\ldots,\mu_r)$ for a weight and $\psi$ and for
the corresponding weight vector, then we have $H_i\psi = \mu_i\psi$.

We rewrite Eq.~(\ref{eq:cartanweyl}) as
\begin{align}
[H_i,E_{ij}]&=(\mu^{(i)}_i-\mu^{(j)}_i)E_{ij},
\end{align}
where the $\mu^{(j)}_i$ are the possible eigenvalues of $H_i$; here
$\alpha_{ij} = \mu_i^{(i)} -\mu_i^{(j)}$.  This shows how the $E_{ij}$
serves as a ``step'' operator that takes an eigenvector with eigenvalue
$\mu_i^{(i)}$ to one with eigenvalue $\mu_i^{(j)}$.

We can write the $H_i$ explicitly in terms of the spectral decomposition:
$$
H_{i}=\sum_{j=1}^d\mu^{(j)}_i\kb{\psi_j}{\psi_j},
$$
where $\ket{\psi_j}$ is the eigenvector corresponding to the $\mu_i^{(j)}$
of $H_i$.

\begin{figure}[t]
\begin{center}
    \resizebox{12cm}{4cm}{\includegraphics{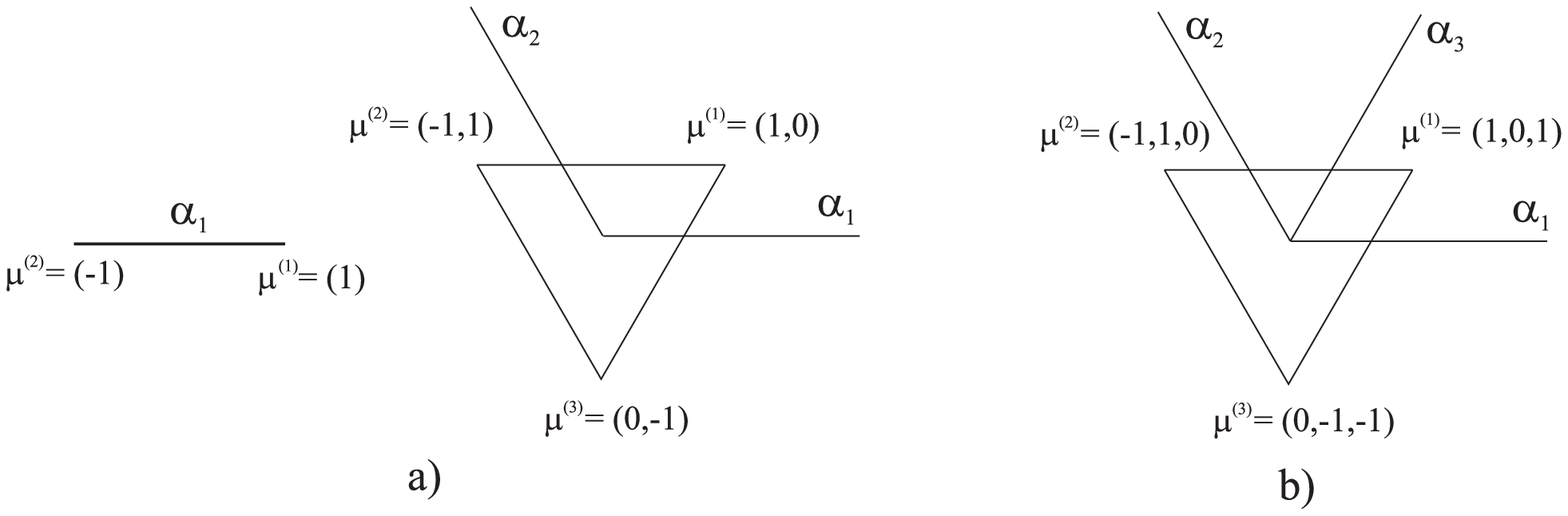}}
    \caption{(a) We illustrate the spaces of roots for $\sltc$ and
      $\slthc$. The basis vectors are simple roots and are indicated by
      $\a_i$. For the weights (vertices) we explicitly write down their
      coordinates in this basis.  We can read off the generators of the
      Cartan subalgebra from the weights. (b) It is sometimes helpful to
      introduce an overcomplete basis. The position of $\a_3$ reflects the
      relation $H_3=H_1+H_2$ for $H_3$ in Eq.~(\ref{eq:H3}) and it is not a
      simple root.}
    \label{fig:GrassX}
\end{center}
\end{figure}
Fig.~\ref{fig:GrassX}a illustrates the situation for $d=2$ and $d=3$.  Hence
for each fundamental representation of $\slnc$ there are $d$ points each
representing an eigenvector $\ket{\psi_j}$.

The important role played by the (fundamental) weights is that they form
the basis for the space of roots and hence they define
the coordinates of the eigenvectors in the space of roots.  The choices of
bases have been made to ensure that all these coordinates are integers.

In the geometric presentation of this data, the different common
eigenvectors of the $H_i$ in the fundamental representation are shown as
points in a lattice.  The step operators $E_{ij}$ define the transitions
between these points.  All the points of the $\slnc$ fundamental
representations are interconnected.  The operator responsible for a
transition from site $\ket{\psi_i}$ to $\ket{\psi_j}$ is the operator
$E_{ij}$ or $E^\dg_{ij}$ for the opposite direction.  It also follows from
Eqs.~(\ref{eq:sl2cinChevalley}) that each segment connecting two
neighboring lattice points corresponds to changing exactly one of the
coordinates of the lattice point by exactly $2$; this just reflects the
fact that the lines correspond to representations of $\sltc$.

The fundamental representations of $\slnc$ algebra contain $r=d-1$
linearly independent $\sltc$ subalgebras each satisfying
Eqs.~(\ref{eq:sl2C}).  However, since the root space diagram is a complete
graph there are in total ${d\choose2}$ linearly dependent $\sltc$
subalgebras corresponding to the number of edges.  The analogues of the
$H_i$, i.e.\ the generators which will be diagonal,
of the ``additional'' $\sltc$ subalgebras are constructed similarly to the
$H_i$'s above.  The only difference is that $1$ and $-1$ on the diagonal are
separated by one or more zeros.  As an example ($d=3$), the remaining
element of the Cartan subalgebra is
\begin{equation}\label{eq:H3}
H_3=\begin{pmatrix}
    1 & 0 & 0 \\
    0 & 0 & 0 \\
    0 & 0 & -1 \\
  \end{pmatrix}.
\end{equation}
Therefore, the rank of the weight vectors equals three and it corresponds to
introducing an overcomplete root basis, see Fig.~\ref{fig:GrassX}b.  Note
that $H_3$ does not correspond to a simple root.  Indeed, the axis $\a_3$ in
Fig.~\ref{fig:GrassX}b can be obtained by a linear combination of $\a_1$
and $\a_2$ which are both positive root vectors.

The fundamental representation of $\slnc$ is a $d$-dimensional
representation with one of the simple roots is assigned $1$ and all others
are zero.  This exactly corresponds to the case where we are looking at the
representation corresponding to $k=1$.  This is a $d$-dimensional space.
The other finite-dimensional representations are obtained by forming tensor
powers of this representation and symmetrizing and antisymmetrizing parts
of the tensor power.  We are looking at the completely symmetric
representations so the combinatorics are particularly simple.  The
completely symmetric representations are obtained by taking the completely
symmetric tensor powers of the fundamental representation.  A well-known
elementary argument shows that the dimension of the space is
${d+k-1\choose d-1} = {d+k-1\choose k}$.


\section{Geometric picture of the completely symmetric representations of the $\slnc$ Lie algebras} \label{app:geom.repn}

For the purpose of this article we are interested in particular
higher-dimensional representations of $\slnc$: the completely symmetric
representations.  The space of roots of $\slnc$ is $(d-1)$-dimensional and
the fundamental root space diagrams are $(d-1)$-simplices.  The $k$-th
lowest completely symmetric representations of $\slnc$ are again
$(d-1)$-simplices.  We can describe their geometric structure as follows:
Each edge connecting a pair of vertices contains $k+1$ equidistant points:
these are completely symmetric states
$\chi_{j_1}=\psi_{(j_2}\dots\psi_{j_d)}$.  The round brackets indicate
symmetrization over the indices inside.

\begin{figure}[h]
\begin{center}
    \resizebox{10cm}{4cm}{\includegraphics{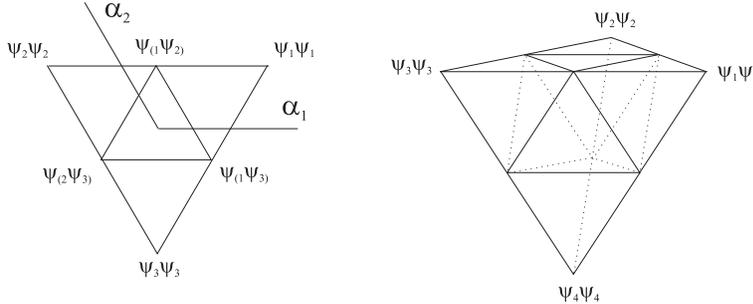}}
    \caption{The space of roots for $d=3$ ($k=2$) on the left and $d=4$
      ($k=2$) on the right. Let us choose the left plot to illustrate the
      explicit decomposition into the $\sltc$ subalgebras. There are two
      lines parallel to each simple root vector. One line (the edge) has
      two segments and it is the second lowest representation of
      $\sltc$. The single-segment line is the fundamental representation of
      $\sltc$. If we direct sum these two algebras and the same thing with
      those corresponding to $\a_2$ they clearly inherit the commutation
      relation of the $\sltc$ algebra. They indeed form the second-lowest
      completely symmetric representation of $\slthc$.}
    \label{fig:sud_unruhX}
\end{center}
\end{figure}

Hence the simplex can be divided into $k$ segments each of length two.  Any
two points are connected provided they lie on a line parallel to an
edge.  This construction determines a lattice.  At each
lattice point lies another completely symmetric state.  The coordinates are given
by a rank $r=d-1$ tensor and they form a higher-dimensional representation
of the $\slnc$ Cartan subalgebra generators.  The states of the $k$-th
lowest completely symmetric representation of $\slnc$ span a ${d+k-1\choose
  k}$-dimensional space.  Two spaces of roots are illustrated in
Fig.~\ref{fig:sud_unruhX}.

All lines of the inner structure connecting the states in the $l$-th
dimensional representations of $\slnc$ determine the $k$-th representations
of $\sltc$ for $l=1\dots k$.  These elementary subalgebras serve as building
blocks for the generators of the given completely symmetric representation
of $\slnc$. The construction is as follows: For a given edge there is a
number of lines parallel to it and so the $\slnc$ subalgebra generators are
formed by a direct sum of these $\sltc$ subalgebras. The reason for a
direct sum is that they, by construction, act on mutually orthogonal
subspaces.
Note that it does not imply that the completely symmetric representations
of $\slnc$ are direct sum representations.  They are actually
irreducible.  The direct sum subalgebras that have been created do not
themselves span mutually orthogonal subspaces, see
Fig.~\ref{fig:sud_unruhX} for illustration.  The consequence is that the
generators constructed in this way manifestly satisfy the commutation
relations for $\slnc$.

For the proof of Lemma~\ref{lem:diagk1} we choose a specific coordinate
system.  The space of roots of $\slnc$ is $(d-1)$-dimensional and we may
choose a (non-orthogonal) coordinate system in many ways.  As illustrated in
Fig.~\ref{fig:GrassX}~(b) for $d=3$ we can choose the axes corresponding to
any pair $\pm\a_i,\pm\a_j$ where $i=j=1\dots3,i\neq j$ and not only the
simple roots $\a_1,\a_2$.  In Lemma~\ref{lem:diagk1} we first choose the
starting highest weight vector (a vertex state) and then it is advantageous
to set up our coordinate system such that the vertex state coordinates are
$(k\dots k)$. As an example, if our vertex state was $\psi_1\psi_1$ in
Fig.~\ref{fig:sud_unruhX} (on the left) the choice of coordinates would be
$\a_1,\a_3$ where $\a_3=\a_1+\a_2$ with the state coordinates being $(22)$.

\bibliography{sud10.bib}

\end{document}

%% file: sud_wiretap1.eepic
\ifx\JPicScale\undefined\def\JPicScale{1}\fi
\unitlength \JPicScale mm
\begin{picture}(107.5,59.38)(0,0)
\linethickness{0.3mm}
\put(24.38,28.12){\line(1,0){9.38}}
\put(24.38,13.12){\line(0,1){15}}
\put(33.75,13.12){\line(0,1){15}}
\put(24.38,13.12){\line(1,0){9.38}}
\linethickness{0.3mm}
\multiput(85.62,28.12)(2.08,0){5}{\line(1,0){1.04}}
\multiput(85.62,13.12)(0,2){8}{\line(0,1){1}}
\multiput(95,13.12)(0,2){8}{\line(0,1){1}}
\multiput(85.62,13.12)(2.08,0){5}{\line(1,0){1.04}}
\linethickness{0.3mm}
\qbezier(33.75,16.25)(34.15,16.02)(38.79,14.38)
\qbezier(38.79,14.38)(43.42,12.75)(47.5,12.5)
\qbezier(47.5,12.5)(53.73,12.83)(59.37,15.48)
\qbezier(59.37,15.48)(65.01,18.12)(71.25,18.75)
\qbezier(71.25,18.75)(74.98,18.68)(78.45,17.35)
\qbezier(78.45,17.35)(81.93,16.02)(85.62,16.25)
\put(85.62,16.25){\vector(1,0){0.12}}
\linethickness{0.3mm}
\qbezier(33.75,23.12)(34.87,22.78)(46.2,22.53)
\qbezier(46.2,22.53)(57.53,22.27)(65.62,26.25)
\qbezier(65.62,26.25)(68.18,27.9)(69.73,30.56)
\qbezier(69.73,30.56)(71.27,33.22)(71.88,36.25)
\qbezier(71.88,36.25)(72.09,38.32)(71.14,40.35)
\qbezier(71.14,40.35)(70.18,42.38)(70.62,44.38)
\qbezier(70.62,44.38)(71.5,47.12)(73.48,49.47)
\qbezier(73.48,49.47)(75.45,51.83)(78.12,51.88)
\qbezier(78.12,51.88)(80.64,51.62)(82.47,49.3)
\qbezier(82.47,49.3)(84.3,46.98)(84.38,44.38)
\qbezier(84.38,44.38)(84.09,41.83)(81.54,40.2)
\qbezier(81.54,40.2)(79,38.58)(77.5,36.25)
\qbezier(77.5,36.25)(75.98,33.51)(74.76,30.46)
\qbezier(74.76,30.46)(73.55,27.42)(75,25)
\qbezier(75,25)(76.6,22.98)(79.58,22.77)
\qbezier(79.58,22.77)(82.56,22.56)(85,23.75)
\put(85,23.75){\vector(2,1){0.12}}
\put(28.75,20.62){\makebox(0,0)[cc]{$\A$}}

\put(90,20){\makebox(0,0)[cc]{$\B$}}

\put(81.25,55){\makebox(0,0)[cc]{$\E^{\otimes n}$}}

\put(50.62,18.12){\makebox(0,0)[cc]{$\N^{\otimes n}$}}

\linethickness{0.3mm}
\multiput(45,26.25)(2.05,0){6}{\line(1,0){1.02}}
\multiput(45,8.12)(0,1.91){10}{\line(0,1){0.95}}
\multiput(56.25,8.12)(0,1.91){10}{\line(0,1){0.95}}
\multiput(45,8.12)(2.05,0){6}{\line(1,0){1.02}}
\linethickness{0.3mm}
\multiput(74.38,59.38)(2.06,0){9}{\line(1,0){1.03}}
\multiput(68.12,44.38)(0.74,1.76){9}{\multiput(0,0)(0.12,0.29){3}{\line(0,1){0.29}}}
\multiput(85.62,44.38)(0.74,1.76){9}{\multiput(0,0)(0.12,0.29){3}{\line(0,1){0.29}}}
\multiput(68.12,44.38)(2.06,0){9}{\line(1,0){1.03}}
\linethickness{0.3mm}
\put(17.5,21.25){\line(1,0){6.88}}
\put(24.38,21.25){\vector(1,0){0.12}}
\linethickness{0.3mm}
\put(95,21.25){\line(1,0){6.88}}
\put(101.88,21.25){\vector(1,0){0.12}}
\linethickness{0.1mm}
\multiput(9.38,41.88)(1.99,0){25}{\line(1,0){0.99}}
\linethickness{0.1mm}
\multiput(58.12,41.88)(1.91,-0.81){9}{\multiput(0,0)(0.32,-0.13){3}{\line(1,0){0.32}}}
\linethickness{0.1mm}
\multiput(74.38,35)(2.01,0){17}{\line(1,0){1}}
\linethickness{0.1mm}
\multiput(62.5,1.25)(0.68,1.93){18}{\multiput(0,0)(0.11,0.32){3}{\line(0,1){0.32}}}
\put(13.75,32.5){\makebox(0,0)[cc]{Alice}}

\put(54.38,53.12){\makebox(0,0)[cc]{Eve}}

\put(105.62,12.5){\makebox(0,0)[cc]{Bob}}

\end{picture}

%% file: sud_codestructure.eepic
\ifx\JPicScale\undefined\def\JPicScale{1}\fi
\unitlength \JPicScale mm
\begin{picture}(125,87.5)(0,0)
\Thicklines
\path(20,60)(30,60)(30,50)(20,50)(20,60)

\Thicklines
\path(50,60)(60,60)(60,50)(50,50)(50,60)

\Thicklines
\path(80,60)(90,60)(90,50)(80,50)(80,60)

\Thicklines
\path(20,55)(15,55)

\Thicklines
\path(15,55)(5,70)

\Thicklines
\path(5,70)(15,85)

\Thicklines
\path(30,52.5)(50,52.5)

\Thicklines
\path(30,57.5)(35,57.5)

\Thicklines
\path(35,57.5)(47.5,75)

\Thicklines
\path(60,52.5)(80,52.5)

\Thicklines
\path(60,57.5)(65,57.5)

\Thicklines
\path(65,57.5)(70,67.5)

\Thicklines
\path(90,52.5)(110,52.5)

\Thicklines
\path(110,52.5)(125,70)

\Thicklines
\path(125,70)(115,85)

\Thicklines
\path(115,85)(15,85)

\Thicklines
\path(90,57.5)(100,57.5)

\Thicklines
\path(100,57.5)(107.5,67.5)

\Thicklines
\path(107.5,67.5)(102.5,75)

\Thicklines
\path(102.5,75)(47.5,75)

\Thicklines
\path(70,67.5)(107.5,67.5)

\put(25,55){\makebox(0,0)[cc]{$V_\A$}}

\put(55,55){\makebox(0,0)[cc]{$U_\N^{\otimes n}$}}

\put(85,55){\makebox(0,0)[cc]{$V_\B$}}

\put(72.5,55){\makebox(0,0)[cc]{$B^n$}}

\put(82.5,70){\makebox(0,0)[cc]{$B_c^n$}}

\put(43.75,82.5){\makebox(0,0)[cc]{$R$}}

\put(13.75,62.5){\makebox(0,0)[cc]{$R'$}}

\Thicklines
\path(20,13.75)(30,13.75)(30,3.75)(20,3.75)(20,13.75)

\Thicklines
\path(50,13.75)(60,13.75)(60,3.75)(50,3.75)(50,13.75)

\Thicklines
\path(30,6.25)(50,6.25)

\Thicklines
\path(20,8.75)(15,8.75)

\Thicklines
\path(15,8.75)(5,23.75)

\Thicklines
\path(5,23.75)(15,38.75)

\Thicklines
\path(30,11.25)(35,11.25)

\Thicklines
\path(35,11.25)(45,26.25)

\Thicklines
\path(60,11.25)(65,11.25)

\Thicklines
\path(65,11.25)(70,18.75)

\Thicklines
\path(70,18.75)(80,18.75)

\Thicklines
\path(80,31.25)(90,31.25)(90,13.75)(80,13.75)(80,31.25)

\Thicklines
\path(15,38.75)(115,38.75)

\Thicklines
\path(45,26.25)(80,26.25)

\Thicklines
\path(115,38.75)(120,32.5)

\Thicklines
\path(90,26.25)(115,26.25)

\Thicklines
\path(115,26.25)(120,32.5)

\Thicklines
\path(90,18.75)(115,18.75)

\Thicklines
\path(60,6.25)(115,6.25)

\Thicklines
\path(115,6.25)(120,12.5)

\Thicklines
\path(120,12.5)(115,18.75)

\put(60,77.5){\makebox(0,0)[cc]{$F$}}

\put(15,15){\makebox(0,0)[cc]{$R'$}}

\put(42.5,36.25){\makebox(0,0)[cc]{$R$}}

\put(60,28.75){\makebox(0,0)[cc]{$F$}}

\put(37.5,55){\makebox(0,0)[cc]{$A^n$}}

\put(37.5,8.75){\makebox(0,0)[cc]{$A^n$}}

\put(25,8.75){\makebox(0,0)[cc]{$V_\A$}}

\put(55,8.75){\makebox(0,0)[cc]{$U_\E^{\otimes n}$}}

\put(85,22.5){\makebox(0,0)[cc]{$V_\D$}}

\put(83.75,8.75){\makebox(0,0)[cc]{$E^n$}}

\put(72.5,21.25){\makebox(0,0)[cc]{$E_c^n$}}

\put(2.5,86.25){\makebox(0,0)[cc]{a)}}

\put(2.5,41.25){\makebox(0,0)[cc]{b)}}

\put(95,60){\makebox(0,0)[cc]{$C_c$}}

\put(105,55){\makebox(0,0)[cc]{$C$}}

\put(97.5,28.75){\makebox(0,0)[cc]{$D$}}

\put(97.5,21.25){\makebox(0,0)[cc]{$D_c$}}

\put(107.5,1.25){\makebox(0,0)[cc]{$\ket{\xi}$}}

\put(100,47.5){\makebox(0,0)[cc]{$\ket{\Omega}$}}

\put(77.5,1.25){\makebox(0,0)[cc]{$\ket{\Psi}$}}

\put(95,60){\makebox(0,0)[cc]{}}

\put(95,60){\makebox(0,0)[cc]{}}

\thicklines
\dashline{1}(100,50)(100,87.5)

\thicklines
\dashline{1}(77.5,3.75)(77.5,41.25)

\thicklines
\dashline{1}(107.5,3.75)(107.5,41.25)

\end{picture}